\documentclass[structabstract]{aa}
\usepackage{txfonts}
\usepackage{amsmath}
\usepackage{natbib}
\bibpunct{(}{)}{;}{a}{}{,} 
\usepackage{graphicx}
\usepackage{threeparttable}
\usepackage{adjustbox}
\usepackage{multirow}
\begin{document}

\title{Near-infrared spectroscopic observations of massive young stellar object candidates in the Central Molecular Zone  \thanks{ Based on observations collected at the European Southern Observatory, Chile, program number 097.C-0208(A)}}  
\author{G. Nandakumar\inst{1}
\and M.~Schultheis\inst{1}
\and   A.~Feldmeier-Krause \inst{2}
\and R.~Sch\"{o}del\inst{3}
\and N.~Neumayer \inst{4}
\and F.~Matteucci \inst{5}
\and N.~Ryde \inst{6}
\and A.~Rojas-Arriagada \inst{7,8}
\and A.~Tej \inst{9}
}

 \institute{ Laboratoire Lagrange, Universit\'e C\^ote d'Azur, Observatoire de la C\^ote d'Azur, CNRS, Blvd de l'Observatoire, F-06304 Nice, France
 e-mail: mathias.schultheis@oca.eu
 \and
 The Department of Astronomy and Astrophysics, The University of Chicago, 5640 S. Ellis Ave., Chicago, IL, 60637, USA
 \and
 Instituto de Astrofísica de Andalucía (CSIC), Glorieta de la Astronomía s/n, 18008 Granada, Spain
  \and
Max-Planck-Institut f\"{u}r Astronomie, K\"{o}nigstuhl 17, 69117 Heidelberg, Germany
\and
Dipartimento di Fisica, Università di Trieste
\and
Lund Observatory, Department of Astronomy and Theoretical Physics, Lund University, Box 43, SE-221 00 Lund, Sweden
\and
Instituto de Astrof\'{i}sica, Facultad de F\'{i}sica, Pontificia Universidad Cat\'olica de Chile, Av. Vicu\~na Mackenna 4860, Santiago, Chile \label{instA1}
\and
Millennium Institute of Astrophysics, Av. Vicu\~{n}a Mackenna 4860, 782-0436 Macul, Santiago, Chile \label{instA2}
\and
Indian Institute of Space Science and Technology, Thiruvananthapuram 695547, India
 }

\titlerunning{Near-infrared spectra of massive YSOs in the Central Molecular Zone}
\authorrunning{Nandakumar et al.}

\abstract  
   { The Central Molecular Zone (CMZ) is a $\sim$\,200 pc region around the Galactic Center. The study of star formation in the central part of the Milky Way is of highest interest as it provides a template for the closest galactic  nuclei.} 
   {We present a spectroscopic follow-up of photometrically-selected young stellar object (YSO) candidates in the CMZ of the Galactic center.  Our goal is to quantify the contamination of this YSO sample by reddened giant stars with circumstellar envelopes and to determine the star formation rate in the CMZ. }  
   {We obtained KMOS low-resolution near-infrared spectra (R\,$\sim$\,4000) between 2.0 and 2.5\,$\mu$m of sources, many of them previously identified, by mid-infrared photometric criteria, as massive YSOs in the Galactic center. Our final sample consists of 91 stars with good signal-to-noise ratio. We separate YSOs from cool late-type stars based on spectral features of CO and $\rm Br{\gamma}$ at 2.3\,$\mu m$ and 2.16\,$\mu m$ respectively. We make use of SED model fits to the observed photometric data points from 1.25 to 24 $\mu m$ in order to estimate approximate masses for the YSOs. }
  { Using the spectroscopically identified YSOs in our sample, we confirm that existing colour-colour diagrams and colour-magnitude diagrams are unable to efficiently separate YSOs and cool late-type stars. In addition, we define a new colour-colour criterion that separates YSOs from cool late-type stars in the H-K$_{\rm S}$ vs H-[8.0] diagram. We use this new criterion to identify YSO candidates in the $|$l$|$ < 1\fdg5, $|$b$|$<0\fdg5 region and use model SED fits to estimate their approximate masses. By assuming an appropriate initial mass function (IMF) and extrapolating the stellar IMF down to lower masses, we determine a star formation rate (SFR) of $\sim$\,0.046\,$\pm$\,0.026\,M$_{\sun}$yr$^{-1}$ assuming an average age of 0.75 $\pm$ 0.25\,Myr for the YSOs. This value is lower than estimates found using the YSO counting method in the literature. }
   {  Our SFR estimate in the CMZ agrees with the previous estimates from different methods and reaffirms that star formation in the CMZ is proceeding at a lower rate than predicted by various star forming models. } 

\maketitle

\section{Introduction}

The CMZ is the innermost $\sim$\,200\,pc region of the Milky Way, covering about $\rm -0\fdg7  < l < 1\fdg8$  in longitude  and $\rm -0\fdg3 < b < 0\fdg2$ in latitude. It is a giant molecular cloud complex with an asymmetric distribution of molecular clouds (see e.g. \citealt{morris1996, Martin2004, oka2005}). The understanding of the physical processes occurring in the CMZ of our Galaxy is crucial for the insight in the formation/evolution of our own Milky Way.

 This prodigious reservoir of molecular gas is in an active region of star formation, with evidence of starburst activity in the last 100,000 years (\citealt{YZ2009}, hereafter YHA09). The  gas pressure and temperature are higher in the CMZ than in the Galactic disk, conditions that favor a larger Jeans mass for star formation and an IMF biased toward more massive stars (see \citealt{Serabyn1996, Fatuzzo2009}). Thus it is essential to understand the modes of star formation and star formation history in the unique environment of the CMZ, both to gain insight into our own Milky Way and to provide a template for circumnuclear star formation in the closest galactic nuclei.

 \cite{2013MNRAS.429..987L} carried out a detailed study of the variations in star formation across the Galactic plane using observational tracers of dense gas (NH$_{3}$(1,1)) as well as star formation activity (masers, HII regions). They showed that there is a deficiency in the star formation activity tracers in the CMZ given the large reservoir of dense gas available. On the other hand, they found that various star formation models predict much higher values of SFR. \cite{2017MNRAS.469.2263B} determined the average SFR across the CMZ using a variety of extra-galactic luminosity--SFR conversion and found it to be comparable to previous measurements made from YSO counting and the free-free emission. Thus they ruled out systematic uncertainties in the SFR measurements as the reason for low star formation in the CMZ. 

The central few parsecs of the Milky Way host a massive young population of stars. There is strong evidence that young stars in the central parsec formed \textit{in situ} (see \citealt{2010RvMP...82.3121G} and references therein). Recent spectroscopic observations have provided further strong evidence to support this. \cite{2015A&A...584A...2F} observed the central >4 pc$^{2}$ of the Galactic centre and identified >100 early-type young stars by spectral classification. They found that early-type stars are centrally concentrated favouring the \textit{in situ} formation of the early-type stars. \cite{2015ApJ...808..106S} mapped a smaller area within 0.28-0.92 pc from SgrA$^{*}$ and found a break in the distribution of young stars at 0.52 pc. They concluded that this break possibly indicated an outer edge to the young stellar cluster in the Galactic center which is expected in the case of \textit{in situ} star formation.
 
 Until recently, most studies of YSOs in the CMZ have been based on infrared photometry  (\citealt{Felli2002, Schuller2006}, YHA09). The YSO phase of a massive star is a relatively brief phase in which they are surrounded by dense envelopes of gas and dust \citep{2007ARA&A..45..481Z}. They are best identified by their point-source infrared radiation as well as excess flux values in the mid-infrared bands. YSOs are classified into three classes/stages depending on their spectral index \citep{1987IAUS..115....1L} as well as spectral energy distribution (SED) models \citep{2006ApJS..167..256R}. \cite{1987ApJ...312..788A} identified Class I YSOs as protostars with infalling envelopes, Class II YSOs as stars with disks and Class III YSOs as those stars having the SEDs of stellar photospheres. Analogous to this classification, \cite{2006ApJS..167..256R} defined three evolutionary stages based on their derived SED model properties : Stage I objects are young protostars embedded in an opaque infalling envelope, Stage II objects are stars surrounded by an opaque disk and dispersed envelope, and Stage III objects are stars with an optically thin disk. \cite{Felli2002} searched for YSO candidates using the mid-infrared excess derived from Infrared Space Observatory (ISO) photometry at 7 and 15 $\mu$m and found  a strong concentration of YSO candidates in the inner Galaxy. \cite{Schuller2006} refined the ISO mid-infrared colour criteria, and argued that slightly extended mid-infrared sources were more likely to be YSOs than point-like mid-infrared sources. YHA09 identified YSO candidates with \textit{Spitzer} photometry at 3.6 -- 24 $\mu$m. Their SED fitting techniques associated most of their YSO candidates  with Stage I objects; they concluded that a recent starburst took place in the CMZ.
 
 The CMZ, however, suffers from very large and spatially-variable interstellar extinction ($\rm A_{V}$ = 20 -- 40 mag, see e.g. \citealt{Schultheis2009}). The significant foreground extinction causes evolved stars with circumstellar envelopes, such as mass-losing asymptotic giant branch (AGB) stars, to have infrared colours similar to those of YSOs. \cite{Schultheis2003} demonstrated that near-infrared spectra are a powerful tool to distinguish YSOs from reddened AGB stars. They found that YSO samples in the CMZ selected by photometric colour criteria are heavily contaminated by AGB stars, red giants and  even supergiants (see their Fig. 5). By contrast, they showed that moderate-resolution spectra in the H and K bands delineate YSOs from evolved stars by the absence  or presence (respectively) of CO absorption at $\sim$ 2.3\,$\mu$m. Detectable YSOs at the distance of the Galactic center ($\sim$8 kpc; see eg: \citealt{2016ApJ...830...17B, 2017ApJ...837...30G}) are all massive, and thus never show 2.3 $\mu$m CO absorption; instead, they are featureless around 2.3 $\mu$m or show \mbox{2.3 $\mu$m} CO in emission (Geballe \& Persson 1987; Carr 1989; Hanson et al. 1997; Bik et al. 2006).

 A recent improvement in YSO selection in the CMZ came from using \textit{Spitzer}/\textit{IRS} spectra to select YSOs. \cite{An2009,An2011}, hereafter An11, presented \textit{Spitzer}/\textit{IRS} 5--35 $\mu$m spectra of 107 YSO candidates selected from 3.6--8.0 $\mu$m \textit{Spitzer} photometry \citep{Ramirez2008}.  An11 identified massive YSOs in the CMZ by the presence of gas-phase absorption from $\rm C_{2}H_{2}$ (13.7 $\mu$m), HCN (14.0 $\mu$m), and $\rm CO_{2}$ (15.0 $\mu$m) as well as strong and broad 15.2 $\mu$m $\rm CO_{2}$ ice absorption. They found that mid-infrared spectra confirm only 33\% of YSO candidates selected by their photometric criteria, and confirm 57\% of YSO candidates selected photometrically by YHA09. \cite{Immer2012} analysed 5--40 $\mu$m \textit{Spitzer}/\textit{IRS} spectra of 57  YSO candidates selected from 7 and 15 $\mu$m ISO colours and spatial extent at 15 $\mu$m \citep{Schuller2006}. They identified 25\% of their sources as YSOs, with an additional 37\% identified as H II regions. There is disagreement in the YSO classification even among the common sources in An11 and \cite{Immer2012} samples, suggesting uncertainties in spectroscopic YSO classification schemes as well. \cite{2015ApJ...799...53K}, using radiative transfer models and realistic synthetic observations, re-examined the YHA09 YSO sample and showed that embedded main sequence stars contaminate the YHA09 sample. These recent studies demonstrate significant contamination of photometrically-selected YSO candidate samples by non-YSOs, which has important implications for CMZ star formation rates derived from photometry (YHA09).

 In this paper, we present moderate-resolution \mbox{2.0--2.5 $\mu$m}  spectroscopic follow-up observations of a sample of 91 photometrically-identified YSO candidates in the CMZ using K-band Multi Object Spectrograph (KMOS, \citealt{2013Msngr.151...21S}) at VLT-UT1 (Antu). Our goal is to distinguish YSOs from evolved late-type stars by their near-infrared spectra. We discuss and show the contaminating evolved late-type stars in different colour-magnitude (CMD) and colour-colour (CCD) diagrams and define a new colour-colour criterion to distinguish them using our spectroscopically identified YSO sample. We estimate the SFR in the CMZ based on YSO counting and on SED fitting techniques.

%
%
%
%

\begin{figure*}[!htbp]
	\centering
		{\includegraphics[width=0.96\textwidth,angle=0]{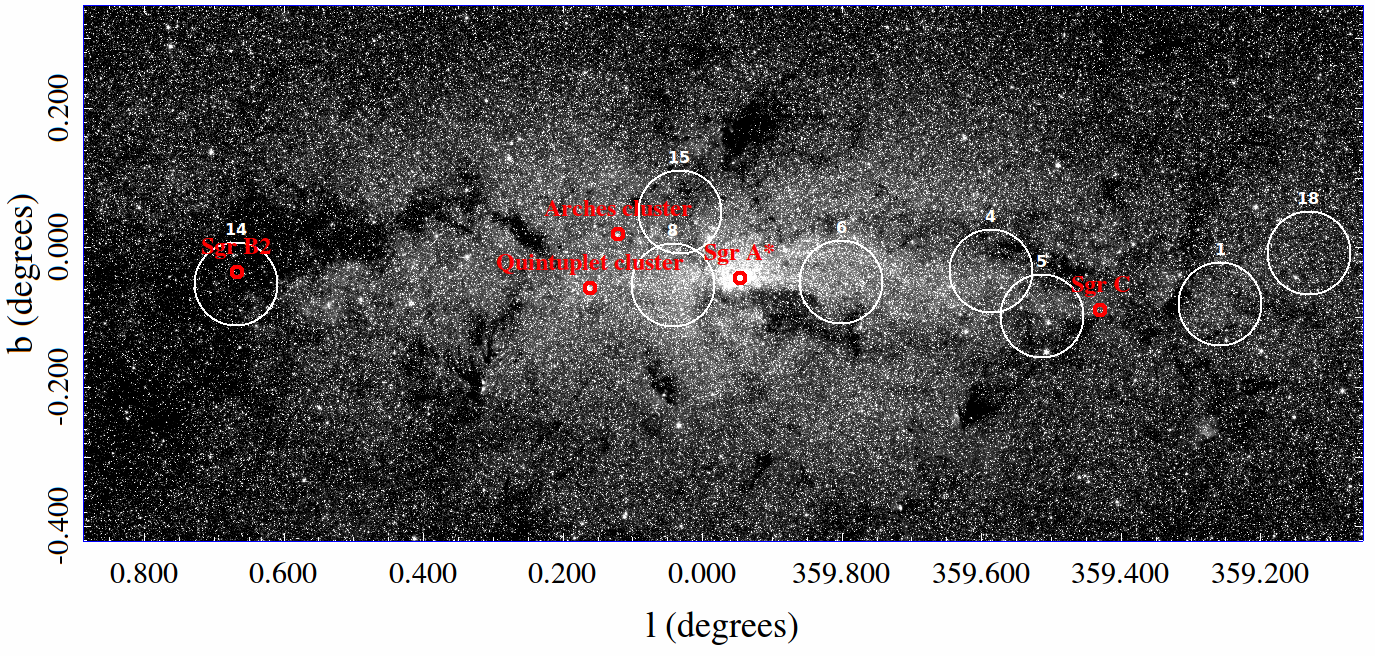}}
		\caption{The field distribution of our observations overlaid on the 3.6 $\mu$m \textit{Spitzer} image \citep{2006JPhCS..54..176S}. The white circles represent the 7\farcm2 diameter fields that have been observed. The numbers are assigned to identify the fields, details of which are given in Table~\ref{OBS_fields}. Shown in red are the locations of prominent sources in the CMZ like Sgr A*, Sgr B2, Sgr C, Quintuplet and Arches clusters. }
\label{planned}
\end{figure*}

\section{Sample selection, observations and data reduction}

\subsection{Sample selection}

We select the sample for our observation from the photometric catalogue of SIRIUS \citep{2006ApJ...638..839N} and point-source catalogue of \textit{Spitzer} IRAC survey of the Galactic center \citep{Ramirez2008}. The JHK$_{\rm S}$ photometry from the SIRIUS catalogue has average 10$\sigma$ limiting magnitudes of J=17.1, H=16.6 and K$_{\rm S}$=15.6 mag, while the 3.6, 4.5, 5.8 and 8.0 $\mu$m bands from the \textit{Spitzer} IRAC catalogue has confusion limits of 12.4, 12.1, 11.7, and 11.2 mag respectively. We divide our sample into three categories of high, medium and low priorities, with the highest priority (priority 1) given to those sources in our sample that are photometrically identified YSO candidates in YHA09. We divide the rest of the sample into medium (priority 2) and low (priority 3) priorities using the following criteria that select sources exhibiting excess emission in mid-infrared regimes :\\
\\
Medium priority \,\,\,\,\,\,\,\,\,\,\,\,\,\,\,\,\,\,\,\,\,\,\,\,\,\,\,\,\,\,\,\,  Low priority\\
$[$3.6$]$-$[$4.5$]$ > 0.5\,\,\,\,\,\,\,\,\,\,\,\,\,\,\,\,\,\,\,\,\,\,\,\,\,\,\,\,\,\,\,\,\,\, [3.6]-[8.0] > 2\\
$[$4.5$]$-$[$5.8$]$ > 0.5\,\,\,\,\,\,\,\,\,\,\,\,\,\,\,\,\,\,\,\,\,\,\,\,\,\,\,\,\,\,\,\,\,\,\,\,K$_{\rm S}$ < 17\\
$[$5.8$]$-$[$8.0$]$ > 1\\
K$_{\rm S}$ < 17\\

KMOS consists of 24 integral field units (IFU) that can be arranged in a 7\farcm2 diameter field per configuration, and it is crucial to prevent the 24 IFUs on the 24 pick-off arms from blocking each other. We use the KMOS ARM Allocator (KARMA) which assigns the maximum number of highest priority targets to the 24 pick-off arms, followed by medium and low priority targets thereby leaving as few arms as possible unallocated. 

\subsection{KMOS observations}

Our spectroscopic observations were carried out with KMOS at VLT-UT1 (Antu) on June 23, 2016. Each of the 24 IFUs in KMOS has a field of view of 2\farcs8$\times$2\farcs8. The spectral resolution of KMOS is R$\sim$4300 with the wavelength range covering \mbox{1.925 $\mu$m - 2.500 $\mu$m}. We prepared 22 fields with unique IFU configuration covering a significant part of the CMZ. But due to bad weather conditions, only 8 fields could be observed with an integration time of 900s for each field in the \textit{nod to sky} mode. The observations were carried out under photometric conditions with seeing $\sim$0\farcs8. The observed field positions are shown in Figure~\ref{planned} and the field details with the number of different priority sources in each field are given in Table~\ref{OBS_fields}.

\begin{table*}[hbt!]
\begin{center}
\caption{Details of the observed fields. The (l,b) of field centers and the number of high, medium, low priority sources, and the manually allocated sky and random sources for the free arms are listed.}
\label{OBS_fields}
\vspace{0.2cm}
\begin{tabular}{c c c c c c c c}
\hline
\hline 
Field No. & l(\degr) & b(\degr) & Priority 1 & Priority 2 & Priority 3 & Sky & Random source \\ 
\hline
1 & 359.2568 &  -0.0813 & 13 & 3 & 3 & 3 & 2 \\ 
 
4 & 359.5854 & -0.0351 & 13 & 5 & 4 & 1 & 1 \\ 

5 & 359.5131 & -0.0986 & 9 & 3 & 11 & 1 & 0 \\ 
 
6 & 359.8000 & -0.0500 & 15 & 1 & 7 & 1 & 0 \\ 

8 & 0.0419 & -0.0547 & 0 & 5 & 18 & 1 & 0 \\ 
 
14 & 0.6674 & -0.0527 & 12 & 9 & 3 & 1 & 0 \\ 
 
15 & 0.0319 & 0.0506 & 3 & 9 & 12 & 2 & 0 \\ 
 
18 & 359.1303 & -0.0080 & 12 & 1 & 5 & 4 & 2 \\ 
\hline 
\hline
\end{tabular} 
\end{center}
 \end{table*}
 
 The fields are in regions with high stellar density as can be seen in the Figure~\ref{planned}. Hence, it was not possible to observe sky in each IFU by dithering or nodding to a new position within the respective field. So we observed a dark cloud G359.94+0.17 ($\alpha$ $\sim$ 266\fdg2, $\delta$ $\sim$ -28\fdg9, \citealt{2001A&A...376..434D}) in free dither mode with 2\farcs8 dither in between two 900s exposures to carry out a proper sky subtraction. The sky observation was carried out after every two field observations. B and A type stars were observed for telluric corrections after every sky offset.

\subsection{Data reduction}

We used the ESO KMOS Recipe Flexible Execution Workbench (Reflex, \citealt{2013A&A...559A..96F}) for data reduction. It organises the science and associated sky and calibration data together based on the calibration source type as well as their proximity in time to science observations. This is followed by dark level correction of frames, flat fielding, wavelength calibration, spatial illumination correction, telluric correction, sky subtraction and cube reconstruction of the science data by dedicated  pipeline recipes (or stages). 

We made use of an IDL routine to remove the Br$\gamma$ absorption line from each telluric spectrum by fitting the Br$\gamma$ line with a Lorentz profile and subtracting it from the telluric spectrum. This routine also removes the stellar continuum by dividing it by a blackbody spectrum. We also removed cosmic rays from the final reconstructed object cube with a 3D version of L.A.Cosmic \citep{2001PASP..113.1420V}.

We extracted spectra from 173 data cubes using \textit{kmos$\_$extract$\_$spec} recipe with the ESO Recipe Execution Tool (EsoRex). \textit{kmos$\_$extract$\_$spec} extracts a spectrum from a data cube with the option of defining a mask manually or automatically by fitting a normalised profile to the image of the data cube. We identify multiple sources in 53 data cubes and extracted their spectra by defining the mask manually. Finally, we extracted nearly 250 spectra and used IRAF median filtering with 7 pixels as filter size to smoothen the spectra. After discarding spectra with a signal-to-noise ratio (S/N) below 20 or having negative flux values, there were 91 spectra left in our sample with good S/N. Among them, there are 15 spectra from data cubes with multiple sources, and we selected the spectrum of the brightest source in the data cube. 

\section{Classification}
In this section, we classify our targets as YSOs and late-type stars based on their spectra. Using this classification, we evaluate different photometric YSO classification criteria and suggest a new criterion to distinguish YSOs and late-type stars.

\subsection{Spectroscopic classification}\label{spectra_classify}

We classify our spectra mainly based on the presence or absence of the $\rm ^{12}CO$ (2,0)-band at 2.3 $\mu m$. CO absorptions bands are typically found in late-type G, K, M giants and AGB stars. In addition, we also use $\rm Br {\gamma}$, found in emission, absorption or with a P-Cygni-type profile in massive YSOs \citep{2013MNRAS.430.1125C}. We have carried out background subtraction during data reduction to make sure that the $\rm Br {\gamma}$ emission lines are intrinsic to the source. Still, we expect contamination from the OB main-sequence/post main-sequence/Wolf Rayet stars, the spectra of some of which also show $\rm Br {\gamma}$ in emission attributed to their stellar wind \citep{Mauerhan2010}. CO band emission at 2.3\,$\mu$m is also considered to be an indication for the presence of a dense circumstellar disk and hence a YSO signature \citep{2006A&A...455..561B}. Some spectra show a featureless continuum at 2.0--2.5 $\mu$m; these could be either YSOs \citep{Greene1996} or dusty late-type carbon-type (WC) Wolf–Rayet stars \citep{Mauerhan2010}.

\begin{figure}[!htbp]
\centering
	{\includegraphics[width=0.49\textwidth,angle=0]{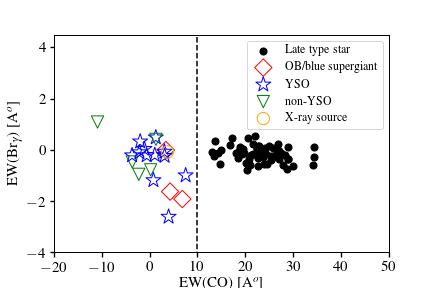}}
		\caption{Equivalent widths measured for the $\rm ^{12}CO$ (2,0) line and $\rm Br{\gamma}$ line for YSOs (blue stars) and late-type stars (black dots). We separate the two populations approximately using the dashed line at EW($\rm Br{\gamma}$) = 10\,\AA\,. Different symbols and colours represent the classification of the SIMBAD matches to our sources by searching within 2\farcs0 (see Section~\ref{spectra_classify}) }
\label{EW}
\end{figure}

\begin{figure}[!htbp]
\centering
\hspace*{-0.2cm}
	{\includegraphics[width=0.55\textwidth,angle=0]{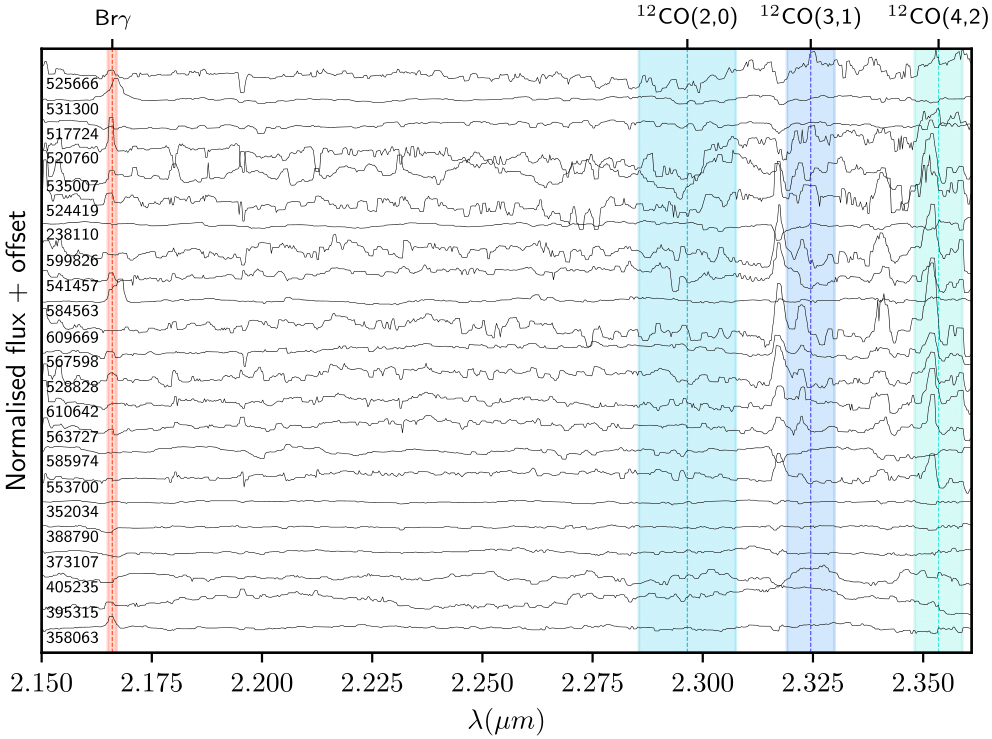}}
		\caption{Normalised spectra of YSOs classified based on the absence of $\rm ^{12}CO$ (2,0) band absorption line and presence of $\rm Br {\gamma}$ emission/absorption line. Dashed lines and shaded areas represent the central wavelengths and range of continuum used for measuring equivalent widths of $\rm Br{\gamma}$, $\rm ^{12}CO$ (2,0), $\rm ^{12}CO$ (3,1) and $\rm ^{12}CO$ (4,2) bands. The SST GC No. for each source is specified adjacent to the corresponding spectra. }
\label{Yso_Spectra}
\end{figure}

\begin{figure}[!htbp]
\centering
\hspace*{-1.1cm}
	{\includegraphics[width=0.55\textwidth,angle=0]{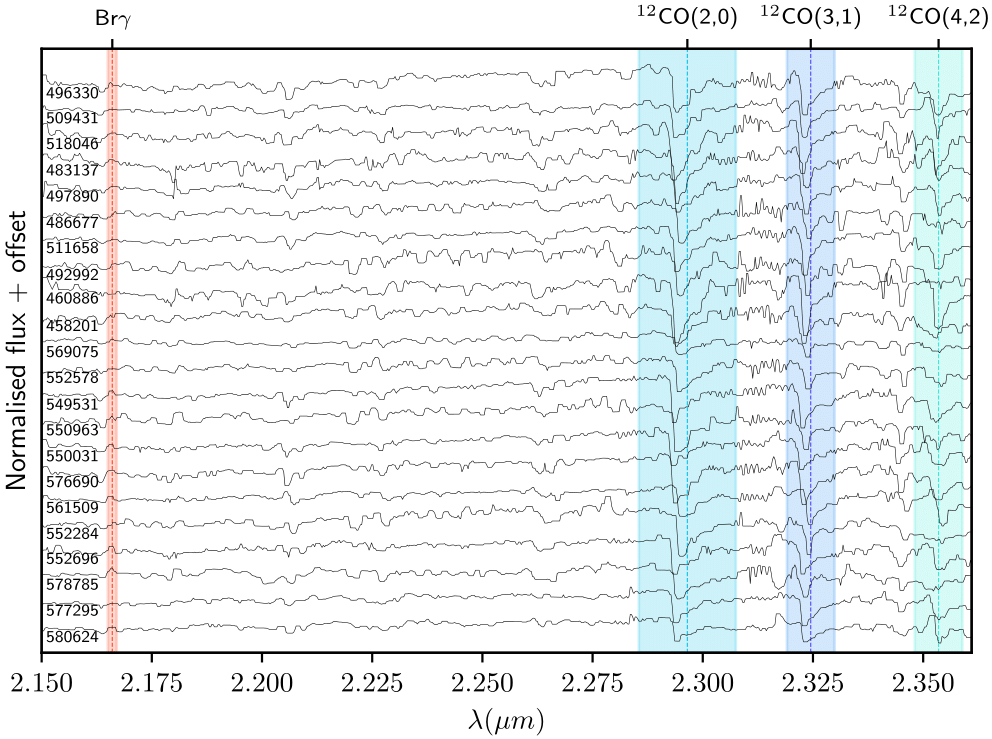}}
		\caption{Same as Fig. 2 but for cool late-type stars in fields 8 and 15, classified based on the presence of $\rm ^{12}CO$ (2,0) band absorption line.}
\label{M_Spectra}
\end{figure}

 We measure the equivalent width (EW) of the $\rm ^{12}CO$ (2,0) band at 2.3\,$\mu$m using the same CO band and continuum points as in \cite{Ramirez2000}. In addition, we measure the EW of $\rm Br{\gamma}$ line at 2.16\,$\mu$m. Figure~\ref{EW} shows the EW(CO) vs EW($\rm Br{\gamma}$) plot of our 91 sources. We find that there are two separate groups of stars with a very evident gap which we approximately mark by the dashed line at EW(CO) = 10 \AA . Positive values of EW indicate that the line is in absorption while negative values indicate it is in emission. Thus we classify the stars to the right of the dashed line as cool, late-type stars and those to the left as YSOs. All cool, late-type stars (represented by filled black circles) lie very close to \mbox{EW($\rm Br{\gamma}$) = 0 \AA}\, indicating the absence of this particular feature in their spectra. The majority of stars we classify as YSOs also show no $\rm Br{\gamma}$ feature, while approximately five stars show $\rm Br{\gamma}$ in emission (EW($\rm Br{\gamma}$)\,<\,1\,\AA\,) and only one star show $\rm Br{\gamma}$ in absorption (EW($\rm Br{\gamma}$)\,$>$\,1\,\AA\,), which also show CO in emission (EW(CO) = -10 \AA).

Based on the above mentioned classification scheme, there are 23 spectroscopically identified YSOs in our sample. Figures~\ref{Yso_Spectra} and\,~\ref{M_Spectra}\, show the reduced normalised spectra of YSOs and cool late-type stars respectively. We searched for previously identified sources in the SIMBAD database with a search radius of 2\farcs0 from each of the 23 YSOs to check their status in the literature. Table~\ref{simbad} lists all 23 sources with their SST (\textit{Spitzer Space Telescope}) GC (Galactic center) No., equatorial coordinates in degrees, distance of the SIMBAD match from the source, source type along with corresponding references and JHK$_{\rm S}$ magnitudes from SIRIUS catalog. There are seven sources in common with An11, out of which only one is confirmed to be YSO and one is considered to be a possible YSO by An11. The remaining five sources have been classified as non-YSOs, with $\#$517724 classified also as a OB super giant star in \cite{2010ApJ...710..706M} based on absorption lines of $\rm Br{\gamma}$ at 2.1661\,$\mu$m, NIII at 2.115 $\mu$m and He I at 2.058, 2.113, and 2.1647\,$\mu$m. Two other sources ($\#$528828 and $\#$599826) have been classified as possible long period variable stars in \cite{matsunaga2009} and \cite{2001MNRAS.321...77G} based on periods estimated using near-infrared observations, though no clear periodicity was found for them. Two sources ($\#$531300 and $\#$584563) have been classified as blue super giant stars based on weak $\rm Br{\gamma}$ emission or absorption feature, NIII and CIV contributions as well as HeI absorption profile at 2.058\,$\mu$m in addition to the Paschen-$\alpha$ (P$\alpha$) excess detected in them \citep{Mauerhan2010}. The counterpart to the source $\#$238110 has been classified as X-ray source in \cite{2003ApJ...599..465M}. Three sources ($\#$358063, $\#$395315 and $\#$520760) are in common with the YHA09 sample, out of which $\#$520760 is classified also as a radio source in \cite{2015MNRAS.446..842D}. The remaining eight sources do not have any counterparts in the SIMBAD database within 2\farcs0.

In Figure~\ref{EW}, we show the sources classified in the literature as X-ray source, OB/blue supergiants and non-YSOs using separate symbols. Regarding the classification of five sources as non-YSOs by An11, the large pixel sizes of \textit{Spitzer/IRS} spectra can lead to mis-identification of sources in high stellar density regions like in the CMZ. Also one non-YSO ($\#$405235) shows $\rm Br{\gamma}$ in absorption and CO in emission, indicating the presence of a dense circumstellar disk and considered to be a massive YSO feature though rarely seen (\citealt{2006A&A...455..561B,2013MNRAS.430.1125C}. Thus we stick with our classification scheme for them. Since no clear periodicity was found for the two sources classified as long period variable stars, we assume them to be YSOs as well. Two blue supergiants ($\#$584563 and $\#$531300) show $\rm Br{\gamma}$ in emission, while no clear emission is seen for $\#$517724 classified as OB supergiant. We made an approximate EW measurement of NIII at 2.115 $\mu$m and found that all three sources mentioned above as well as the radio source $\#$238110 show clear emission feature, not seen in the rest of our YSOs. Thus we consider their classifcation as supergiants to be acceptable. Based on these measurements we conclude that none of the rest of our YSOs is a O/B supergiant.  

 \begin{figure*}[!htbp]
 
	\includegraphics[width=.35\textwidth]{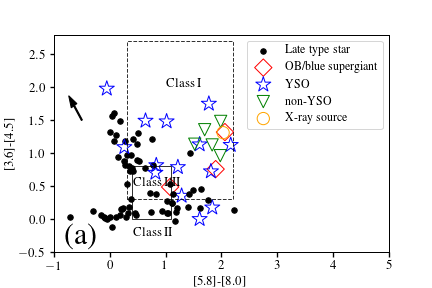}
	\quad \includegraphics[width=.35\textwidth]{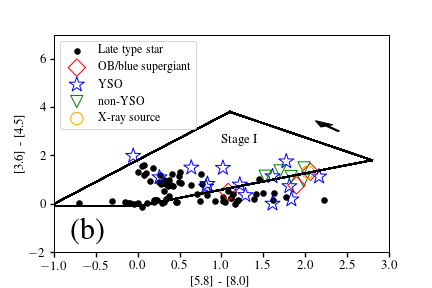}
	\quad \includegraphics[width=.30\textwidth]{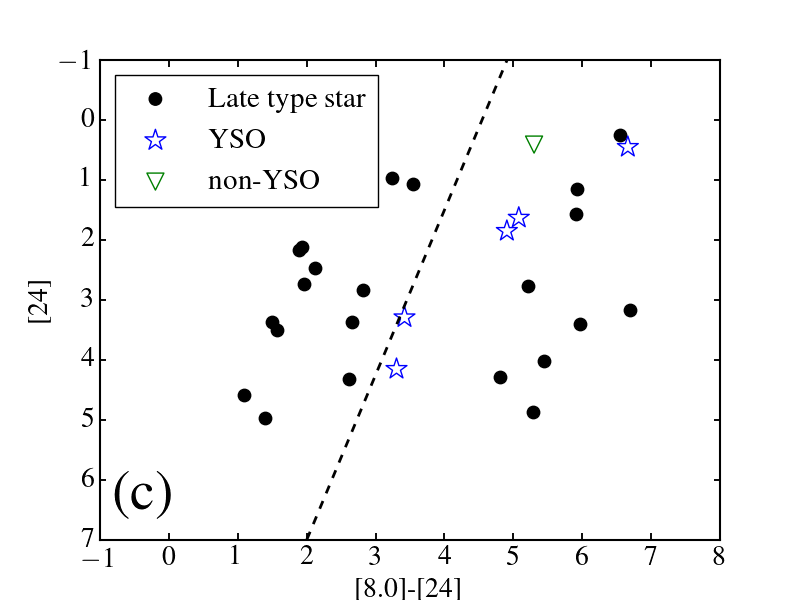}
		\caption{(a) : [5.8]-[8.0] vs [3.6]-[4.5] diagram used to identify different classes of YSOs based on the disk and envelope models of low mass YSOs as shown in \cite{2004ApJS..154..363A} and \cite{2004ApJS..154..367M}. Class II YSOs are expected to be concentrated in the small box, while Class I YSOs in the bigger one. (b) : Same colour-colour diagram as (a) with the region shown by black polygon where Stage I YSOs are expected to lie \citep{2006ApJS..167..256R}. (c) : [8.0]-[24] vs [24] diagram showing the criteria used by YHA09 to choose their sample of possible YSO candidates (region to the right of the dashed line). The black arrows in (a) and (b) represent the extinction vector estimated for A$_{K}$ = 2\,mag (typical of the CMZ) using the $\frac{A_{\lambda}}{A_{K}}$ relations from \cite{2009ApJ...696.1407N}. In each diagram, the spectroscopically identified YSOs are shown using blue star symbols and the cool late-type stars using black filled circles. Other symbols and colours represent the classification of the SIMBAD matches to our sources by searching within 2\farcs0 (see Section 3.1) }
\label{CC_lit}
\end{figure*}

\begin{table*}[hbt!]
\begin{center}
\begin{threeparttable}[b]
\caption{Details of the search for previously identified sources within 2\farcs0 of our spectroscopically identified YSOs in the SIMBAD database. For each source represented by their SST GC No., their equatorial coordinates in degrees, distance from the source, source type along with corresponding references and JHK$_{\rm S}$ magnitudes from SIRIUS catalog are listed. For sources with no counterpart within 2\farcs0 and no valid photometry in J or H bands, we use '...' for the corresponding column. } 
\label{simbad}
\vspace{0.2cm}
\begin{tabular}{c c c c c c c c c c}
\hline
\hline 
SST GC No. & RA (\degr) & DEC (\degr) & Distance (in \arcsec) & Source type & J (mag) & H (mag) & K$_{\rm S}$ (mag) & Field\\ 

\hline
238110 & 265.93584 & -29.67287 & 0.64 & X-ray source\tnote{a} & 17.19 &  13.64 & 11.84  & 18 \\ 
 
517724 & 266.40542 & -28.89823 & 0.07 & OB supergiant\tnote{b}, non-YSO\tnote{c} & 15.75 & 12.80 & 11.23 & 15 \\ 
 
520760 & 266.41002 & -28.89086 & 1.54 & radio source\tnote{d}, YSO\tnote{e} & ... & 15.59 & 13.17 & 15 \\

524419 & 266.41569 & -28.89559 & 0.50 & non-YSO\tnote{c} & ... & 15.86 &  14.02 & 15  \\  
 
525666 & 266.41759 & -28.89117 & 0.04 & non-YSO\tnote{c} & ... & 14.49 &  12.74  & 15 \\ 

531300 & 266.42633 & -28.87979 & 0.17 & Blue super giant\tnote{b} & 14.70 & 11.67 & 10.11 & 15 \\ 

535007 & 266.43185 & -28.87358 & 0.05 & non-YSO\tnote{c} & ... & 16.05 &  13.81 & 15\\ 
 
528828 & 266.42248 & -28.90728 & 0.11 & Long period variable star\tnote{f} & ... & ... & 13.69 & 8 \\ 

541457 & 266.44147 & -28.90716 & ... & ... & ... & 14.28 & 12.53 & 8\\ 

553700 & 266.45999 & -28.91312 & ... & ... & 15.11 & 13.09 & 12.27 & 8\\ 

563727 & 266.47539 & -28.97628 & ... &  ... & ... & 14.75 & 12.97 & 8 \\ 

567598 & 266.48123 & -28.93786 & ... & ... & ... & 14.71 & 12.92 & 8\\ 

609669 & 266.54470 & -28.92073 & ... &  ... & ... & 16.24 & 14.17 & 8\\ 

610642 & 266.54618 & -28.92802 & 0.05 &  Maybe YSO\tnote{c} & ... & 15.26  &  12.69 & 8\\ 

584563 & 266.50690 & -28.92095 & 0.17 &  Blue super giant\tnote{b} & 13.45 & 10.69 & 9.12 & 8\\ 

585974 & 266.50900 & -28.95653 & ... & ... & 16.80 & 11.73 & 9.42 & 8\\ 

599826 & 266.52982 & -28.93144 & 0.52 & Long period variable star\tnote{f} & ... & ... & 13.87 & 8\\ 

373107 & 266.17890 & -29.33702 & ... & ... & 16.48 & 13.59 & 11.94 & 4\\ 

388790 & 266.20389 & -29.39521 & 0.40 &  non-YSO\tnote{c} & ... & 14.12 & 12.05 & 5\\ 

352034 & 266.14518 & -29.39360 & ... & ... & ... & 12.52 & 10.95 & 5\\ 

395315 & 266.21438 & -29.34097 & 0.02 & YSO\tnote{e} & ... & 15.95 & 13.65 & 4\\ 

405235 & 266.23023 & -29.26057 & 0.28 & non-YSO\tnote{c} & ... & 14.77 & 12.97 & 4\\ 

358063 & 266.15475 & -29.31017 & 0.14 & YSO\tnote{e} & 15.59 &  13.42 & 11.79 & 4\\ 

\hline 
\hline
\end{tabular} 
\begin{tablenotes}
     \item[a] \citealt{2003ApJ...599..465M}
     \item[b] \citealt{Mauerhan2010}
     \item[c] \citealt{An2011}
     \item[d] \citealt{2015MNRAS.446..842D}
     \item[e] \citealt{YZ2009}
     \item[f] \citealt{matsunaga2009, 2001MNRAS.321...77G}
    \end{tablenotes}
\end{threeparttable}  
\end{center}
 \end{table*}

\subsection{Classification using photometric criteria}\label{hkh8}

Several previous studies have made use of colour-colour diagrams (CCDs) to define criteria in order to classify YSOs. We are trying with our spectroscopic sample of YSOs and non-YSOs to establish  new and more reliable photometric criteria in order to distinguish YSOs from non-YSOs. Initially, we make use of some YSO classification criteria using CCDs implemented in the literature to check if they are able to classify and separate our sample of spectroscopically identified YSOs from late-type stars. For this, we obtain the photometry of point sources at 3.6\,$\mu$m, 4.5\,$\mu$m, 5.8\,$\mu$m, 8.0\,$\mu$m \citep{Ramirez2008}, and 24\,$\mu$m from the MIPSGAL catalog \citep{2015AJ....149...64G}, and we merge these with our data by searching within radii of 2\farcs0. We find 87 sources with valid 3.6 to 8.0\,$\mu$m photometry and 28 sources with valid 24\,$\mu$m photometry in our sample. 

Figure~\ref{CC_lit} shows two CCDs and a colour-magnitude diagram (CMD) with respective criteria from the literature to classify YSOs. In Figure~\ref{CC_lit}(a), we plot [5.8]-[8.0] vs [3.6]-[4.5] with a small and big box representing the regions belonging to Class II and Class I YSOs respectively. Those areas come from the disk and envelope models of low mass YSOs as shown in \cite{2004ApJS..154..363A} and \cite{2004ApJS..154..367M}. Figure~\ref{CC_lit}(b) shows the same plot, but with a polygon used to define the region enclosing Stage I YSOs as defined in \cite{2006ApJS..167..256R}. YHA09 used the CMD ([8.0]-[24] vs [24]) in Figure~\ref{CC_lit}(c) to choose their sample of possible YSO candidates in the CMZ by considering all sources lying to the right of the dashed line as YSOs. We estimate and show the extinction vector (black arrow) for the two CCDs, assuming an A$_{K}$ of 2 mag (corresponding to A$_{V}$=30 mag; typical of the CMZ from \cite{Schultheis2009}) using the $\frac{A_{\lambda}}{A_{K}}$ relations from \cite{2009ApJ...696.1407N}. It is clear from each plot that, even after taking the extinction into account, there is severe contamination from the late-type stars in the regions defined to contain YSOs and that there is no clean colour-colour criterion visible.

\begin{figure}[!htbp]
\centering
	{\includegraphics[width=0.49\textwidth,angle=0]{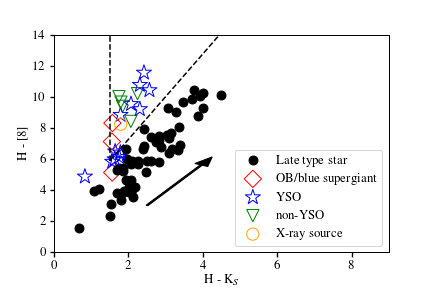}}
		\caption{H-K$_{S}$ vs H-[8.0] diagram that shows a clear trend for cool late-type stars separating the YSOs in our sample. The dashed line represents our proposed criterion to separate YSOs and late-type stars. The black arrow represent the extinction vector estimated for A$_{K}$ = 2\,mag (typical of the CMZ) using the $\frac{A_{\lambda}}{A_{K}}$ relations from \cite{2009ApJ...696.1407N}.}
\label{CC_our}
\end{figure}

Hence, we test different combinations of colours and magnitudes in order to clearly separate YSOs from cool late-type stars in our sample. Only in the CCD, H-K$_{\rm S}$ vs H-[8.0], as shown in Figure~\ref{CC_our}, we see a clear linear trend followed by the late-type stars, while the YSOs exhibit redder H-[8.0] colours and are thus clearly separated. We define a rough criterion to separate them, taking into account the H-K$_{\rm S}$ cut at 1.5 estimated to remove foreground sources (assuming A$_{V}$\,=\,30\,mag  and using extinction laws of \cite{2009ApJ...696.1407N}). A similar H-K$_{\rm S}$ cut was also suggested by \cite{2010A&A...511A..18S} to remove foreground sources based on their study toward the central parsec of the Galaxy. Here again, we estimate the extinction vector as mentioned before and show that it is almost parallel to the line separating YSOs from late-type stars. Thus the extinction does not greatly affect our criterion. We want to stress that our sample is small and this criterion needs to be confirmed by a larger sample:
   
\begin{equation}
(H-[8.0]) = 2.75\times(H-K_{\rm S}) + 1.75 ; \,\,1.5 < (H-K_{\rm S}) \leq 5 
\label{eq1}
\end{equation}  

\section{Mass estimates using SED model fits }

In the above section, we spectroscopically identified 23 YSOs from among 91 sources in the CMZ. In this section, we construct their near and mid-infrared SEDs and fit them using synthetic SED models for YSOs to constrain their stellar parameters (e.g. stellar radius R$_\star$, effective temperature T$_{\rm eff}$, stellar mass M$_{\star}$, total stellar luminosity L$_{\star}$, extinction in V-band A$_{V}$).

\subsection{Robitaille et al. models}

\cite{2006ApJS..167..256R} presented a set of approximately 20,000 radiative transfer models with corresponding SEDs assuming an accretion scenario with a central star surrounded by an accretion disk, infalling envelope and bipolar cavities, i.e., YSOs. \cite{2007ApJS..169..328R} presented a tool to fit these YSO model SEDs to observations, providing a range in the parameter (e.g. stellar mass, total luminosity, extinction in V-band, envelope accretion rate, age) space corresponding to a set of best fit models. 

These models are largely in use to estimate approximate values of stellar parameters for a photometrically or spectroscopically identified sample of YSOs. YHA09 and An11 used these models in order to classify YSOs into different evolutionary stages based on the envelope infall rate and disk accretion rate of each source. YHA09 in turn estimate the star formation rate (SFR) in the CMZ using masses they obtain from the SED fits.

Recently, \cite{2017A&A...600A..11R}, hereafter R17, introduced an improved set of SED models for YSOs that covers a much wider range of parameter space and excluding most of model-dependent parameters in addition to several other improvements. Unlike previous models, there are 18 different sets of models with increasing complexity that varies from a single central star to a star in an ambient medium surrounded by accreting disk, infalling envelope and bipolar cavities as described in detail in R17. We use the latest R17 models to fit the SEDs of sources in our sample to estimate stellar parameter such as stellar radius, luminosity or effective temperature. We will use these parameters to determine the stellar masses of our YSO sample.

\subsection{SED fits}

We construct the SEDs for our sample using wavelengths ranging from 1.25 - 24\,$\mu$m. As mentioned in section~\ref{hkh8}, we use the JHK$_{\rm S}$ photometry from the SIRIUS catalogue, 3.6 to 8.0\,$\mu$m photometry from \cite{Ramirez2008} and 24\,$\mu$m photometry from \cite{2015AJ....149...64G}. In addition, we use the 15\,$\mu$m photometry from the ISOGAL point source catalog (\citealt{2003A&A...403..975O,2003yCat.2243....0O}) so that we constrain the SEDs over a large wavelength range. Searching within 2\farcs0 of YSOs in our sample, we find seven sources in the ISOGAL PSC out of which only two have valid 15\,$\mu$m magnitudes. Within the same search radius, we find six YSOs with a match in the 24 $\mu$m catalogue, all of which have valid photometry.

Thus, among 23 spectroscopically identified YSOs in our sample, in addition to 1.25 - 8.0\,$\mu$m magnitudes, six sources have only 24\,$\mu$m magnitudes, two sources have only 15\,$\mu$m magnitudes and there are no sources with valid magnitude determined at both 15\,$\mu$m as well as 24\,$\mu$m. We find that the SED fitting by the model requires data points at $\lambda$>12 $\mu$m to give reliable results. For that reason, we carried out SED fits using the SED fitting tool only for these eight sources using the above mentioned set of photometry.

\begin{figure*}[!htbp]
	\hspace*{-1.1cm}\includegraphics[width=.35\textwidth]{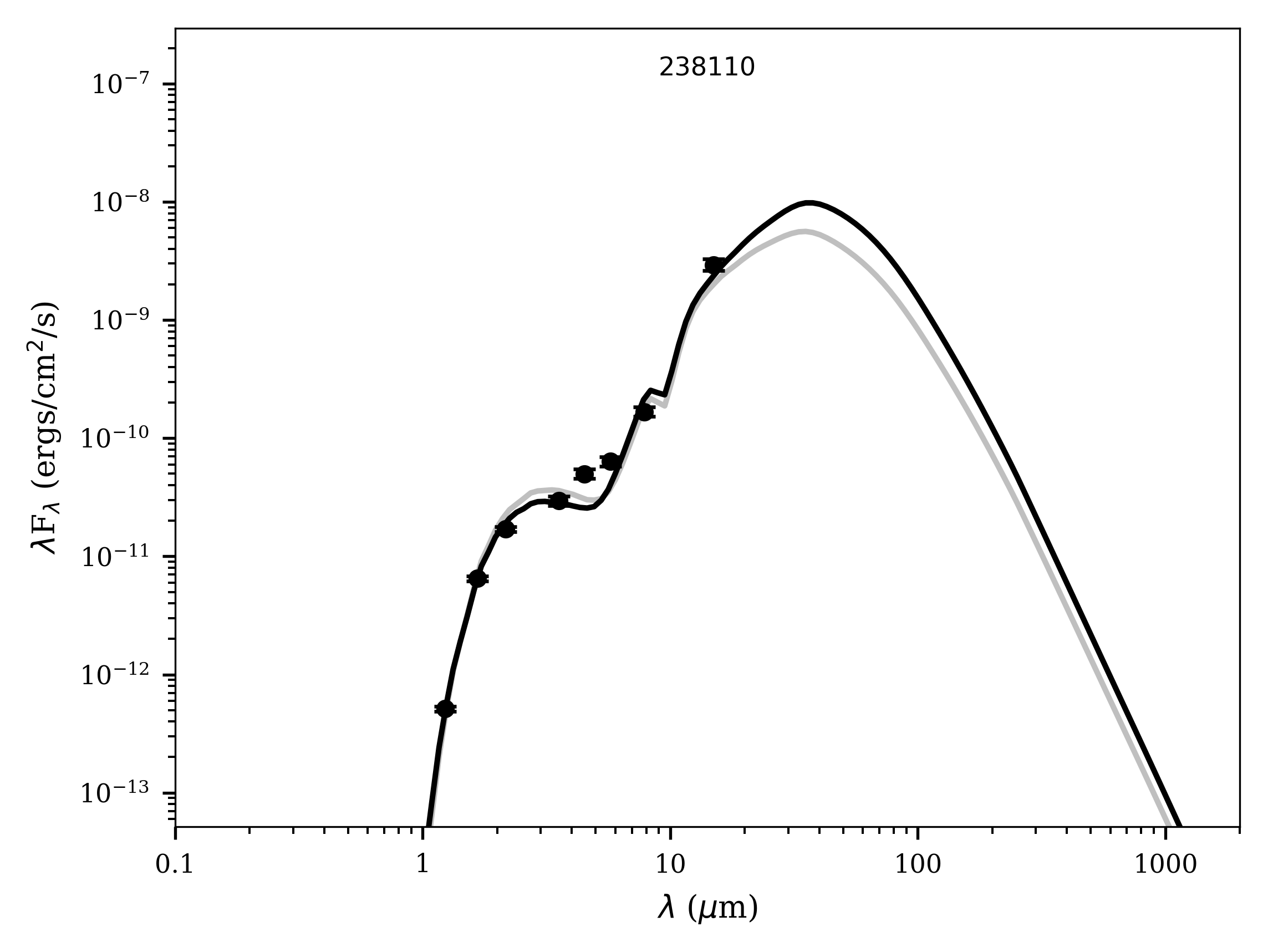}
	\quad \includegraphics[width=.35\textwidth]{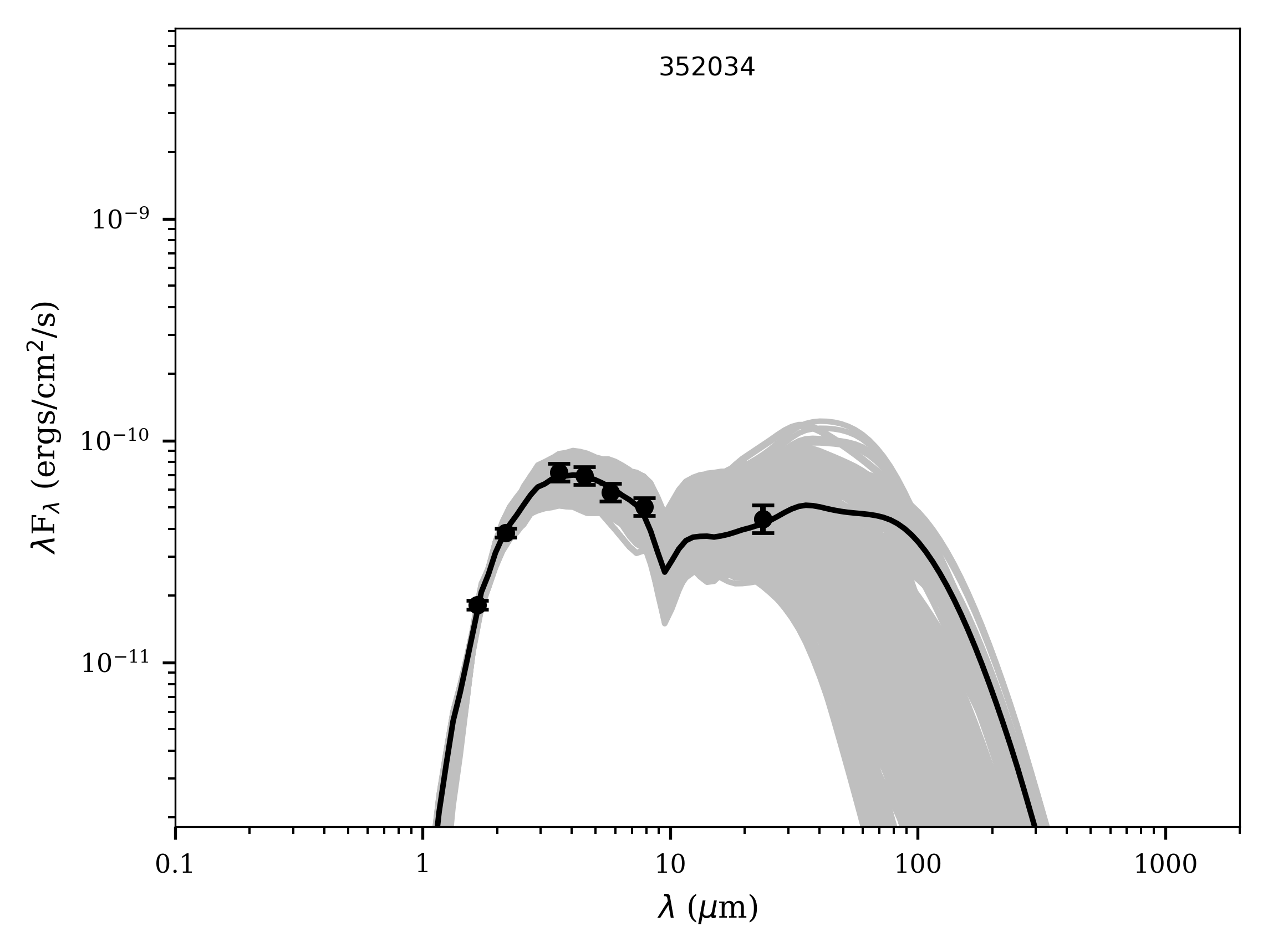}
	\quad \includegraphics[width=.35\textwidth]{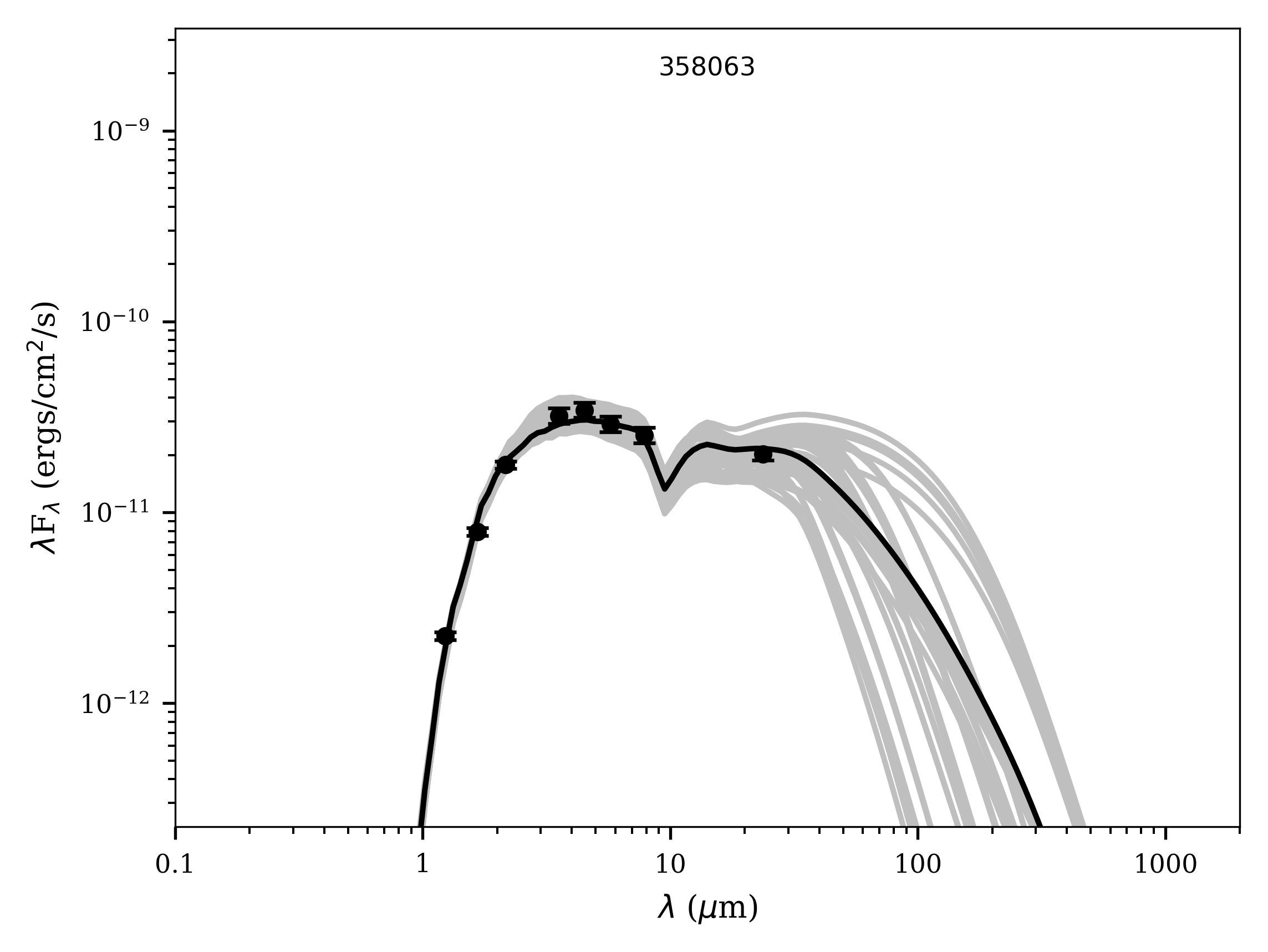}
	\hspace*{-1.1cm}\quad \includegraphics[width=.35\textwidth]{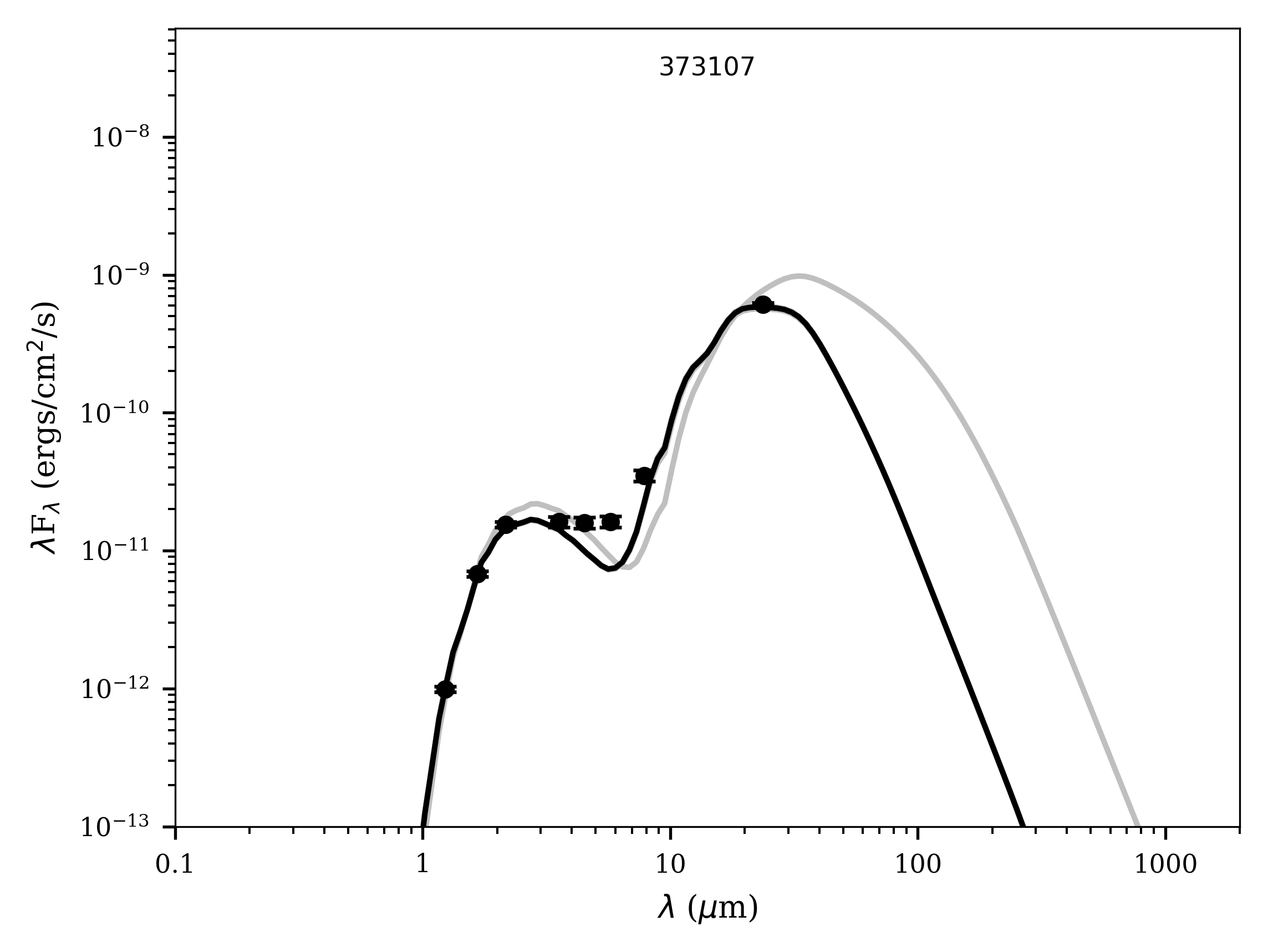}
	\quad \includegraphics[width=.35\textwidth]{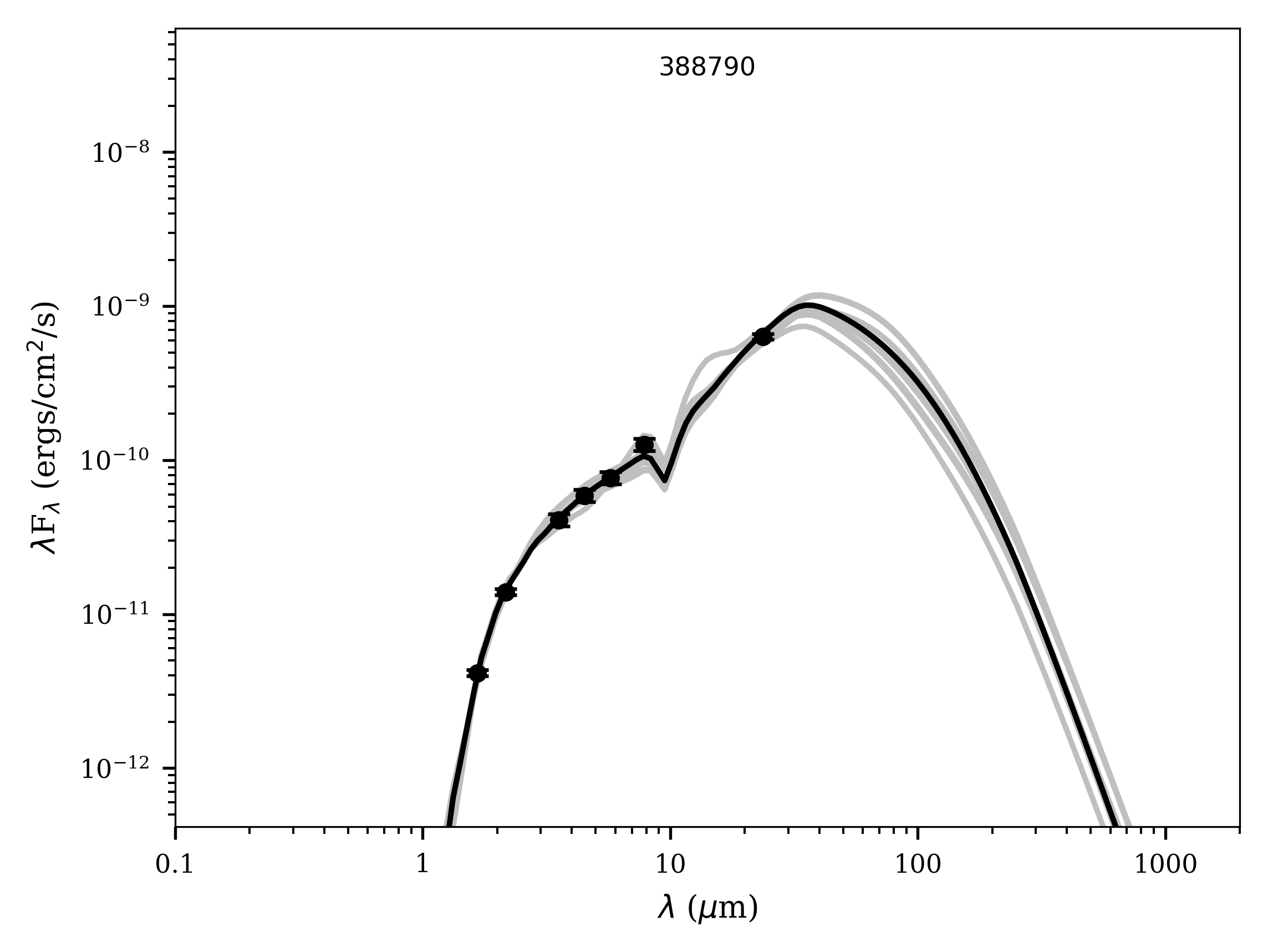}
	\quad \includegraphics[width=.35\textwidth]{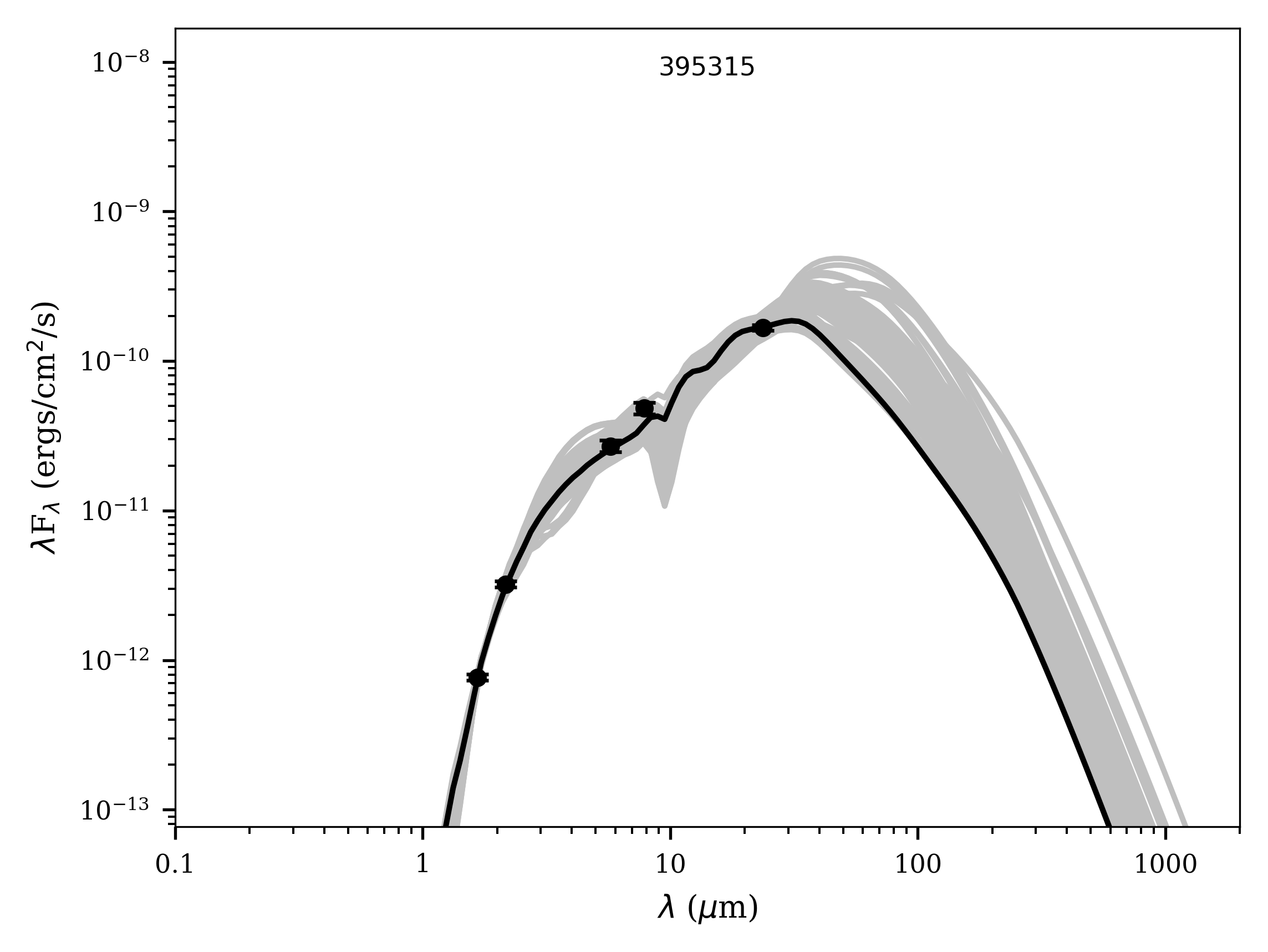}
	\hspace*{1.8cm}\quad \includegraphics[width=.35\textwidth]{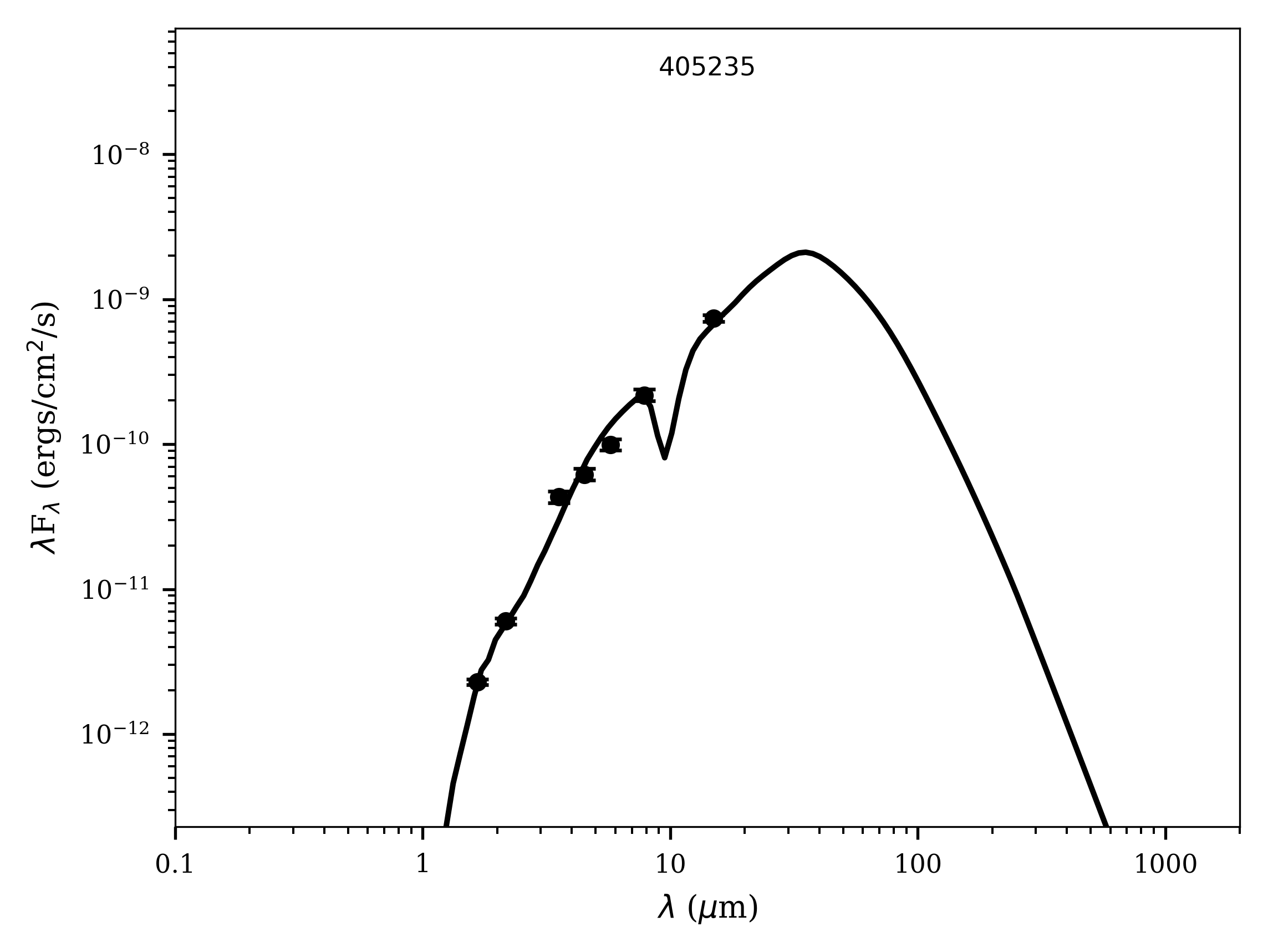}
	\quad \includegraphics[width=.35\textwidth]{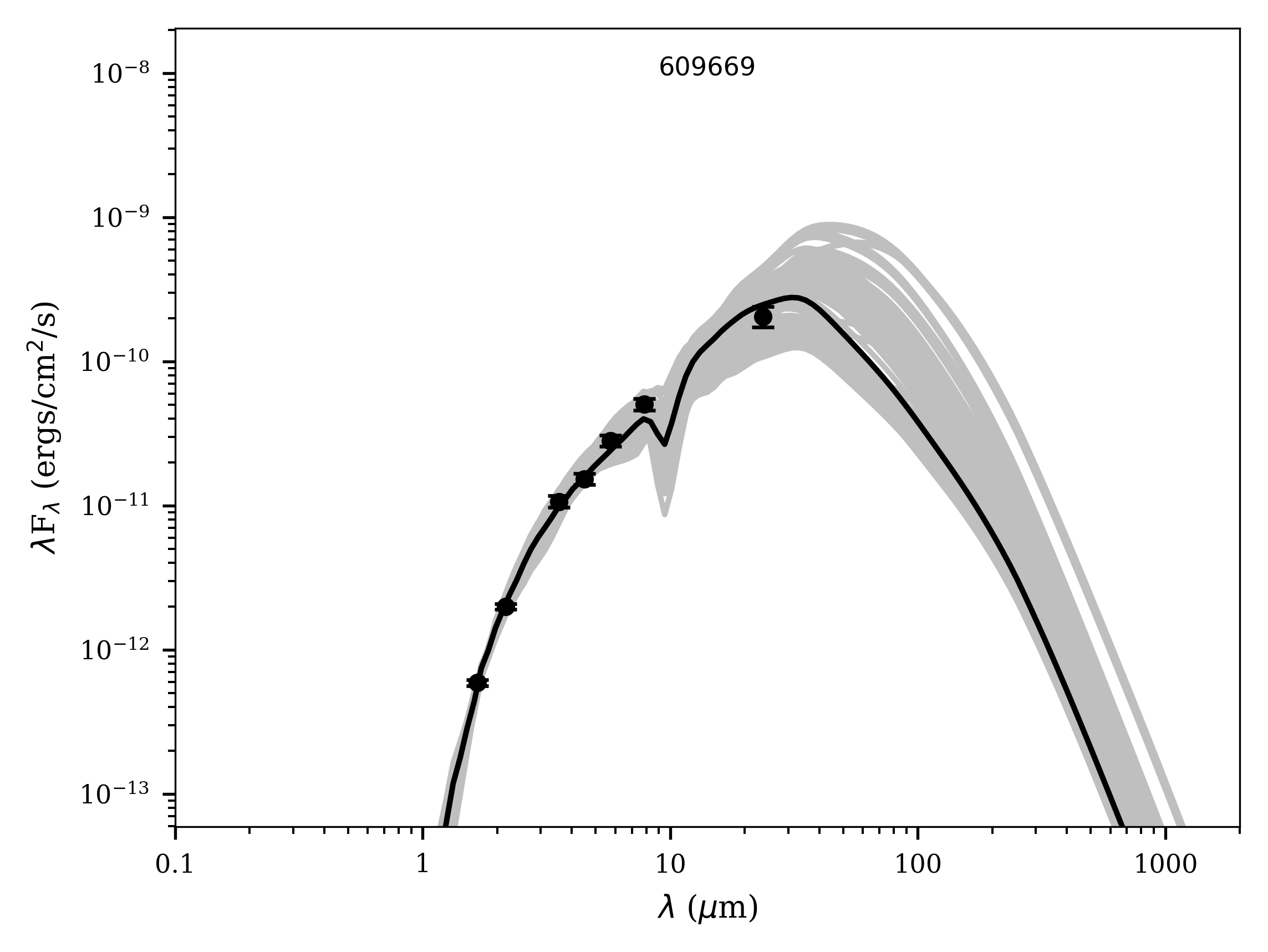}
		\caption{SED fits for 8 YSOs in our sample with available photometric data upto 24 $\mu$m, performed by the SED fitter tool using different YSO SED models in R17. The best fit is shown using solid black line and gray lines show all other fits that satisfy the criteria : $\chi^{2}$ - $\chi_{\rm best}^{2}$ $<$ 5 per data point.  }
\label{sedfits}
\end{figure*}

 We assume the source distance to be in the range 7 kpc $<$ R $<$ 9 kpc from the Sun and interstellar extinction along the line of sight to the Galactic center to be in the range 20 mag $<$ A$_{V}$ $<$ 50 mag \citep{Schultheis2009}, ensuring that these sources belong to the CMZ. These assumptions ensure that the conditions at the Galactic center are considered while fitting the model SEDs to our observed SEDs. We assume typical errors of 0.05\,mag for JHK$_{\rm S}$ photometry and 0.1 mag for 3.6 to 8.0\,$\mu$m  photometry, while ISOGAL and MIPSGAL catalogues provide typical errors of $\sim$0.05 mag for 15\,$\mu$m and 0.1 mag for 24\,$\mu$m photometry respectively. To include reasonable fitting results, we select all SEDs that satisfy $\chi^{2}$ - $\chi_{\rm best}^{2}$ $<$ 5 per data points for each source in all 18 model sets. $\chi_{\rm best}^{2}$ represents $\chi^{2}$ value of the best fit for each model set. Thus each source SED is fitted with 18 different model sets, each of which gives a best fit SED with a $\chi_{\rm best}^{2}$ value. For each source, we select the model set corresponding to the best fit SED with the lowest $\chi_{\rm best}^{2}$ value as the one that best represents the evolutionary stage of the source. Figure~\ref{sedfits} displays typical examples of SED fitting results for eight YSOs in our sample. 
 
\subsection{Fit parameters and mass}\label{mass}
 
 We choose the model set corresponding to the best fit SED with the lowest $\chi_{\rm best}^{2}$ value as mentioned above and estimate mean values of A$_{V}$, T$_{\rm eff}$ and stellar radius, R$_{\star}$ from all the fits satisfying the $\chi^{2}$ cut in the chosen model set. We estimate approximate values for the stellar luminosity, L$_{\star}$, using the Stefan-Boltzmann law from T$_{\rm eff}$ and R$_{\star}$, assuming solar T$_{\rm eff}$ to be 5772 K. To determine an approximate mass for each YSO, we use the pre-main sequence (PMS) tracks for stars with metallicity of Z = 0.02 and mass range of 0.8 M$_{\sun}$ to 60 M$_{\sun}$ from \cite{1996A&A...307..829B}. The masses are sampled in a non-uniform manner with stellar tracks provided for 0.8, 1.0, 1.5, 2.0, 3.0, 5.0, 9.0, 15.0, 25.0 and 60.0 M$_{\sun}$.
 
\begin{figure*}[!htbp]
\centering
	{\includegraphics[width=0.56\textwidth,angle=0]{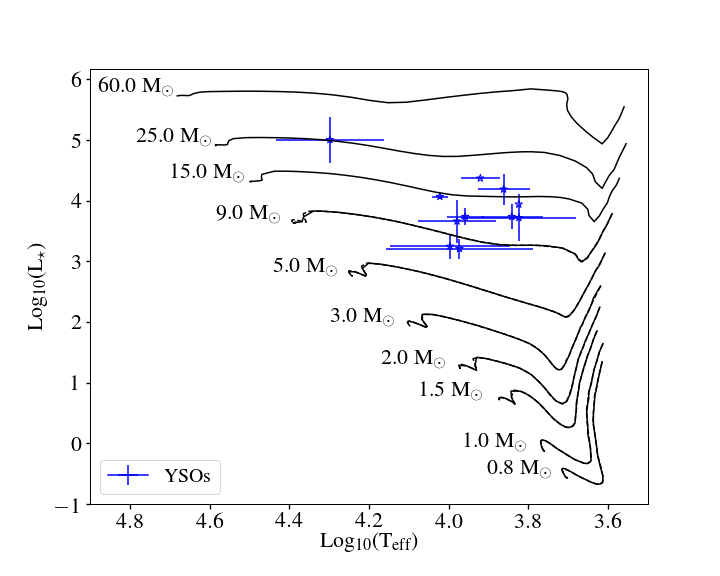}}
		\caption{log$_{10}$(L$_{\star}$) vs log$_{10}$(T$_{\rm eff}$) diagram showing the location of YSOs (blue stars) and the pre-main sequence (PMS) stellar tracks from \cite{1996A&A...307..829B} for different masses in black. The errors in T$_{\rm eff}$ and L$_{\star}$ are also shown.}
\label{LT}
\end{figure*}

We calculate the separation of each source from the stellar track for each mass in the log$_{10}$(L$_{\star}$) - log$_{10}$(T$_{\rm eff}$) space, and assign them the mass corresponding to the track at the least separation. This exercise is carried out for all fits that satisfy the $\chi^{2}$ cut in the chosen model set and we estimate a mean mass from them. The standard deviation in mass from all SED fits can be used to make a rough estimate of the error. A zero error for mass is obtained when there is only one SED fit or when there is a single closest stellar track to source positions from all SED fits of the chosen model set. We assume their mass uncertainties to be limited by the mass sampling of PMS tracks. Figure~\ref{LT} shows the log$_{10}$(L$_{\star}$) vs log$_{10}$(T$_{\rm eff}$) plot with the stellar tracks and location of 8 spectroscopically identified YSOs. Mass estimates range from $\sim$ 8 to 20 M$_{\sun}$, as expected for high mass YSOs. 

We estimate the uncertainties for T$_{\rm eff}$, L$_{\star}$ and A$_{V}$ similarly from standard deviation in their values from all SED fits of the chosen model set. Table~\ref{fit_param} lists the main model fit parameters and the estimated masses for the YSOs. The A$_{V}$ values estimated by the SED models are mostly close to the lower end of our constraints (20 mag\,$<$\,A$_{V}$\,$<$\,50 mag), which is not the expected case. So we used the extinction map of \cite{Schultheis2009} to estimate the foreground visual extinction close to the location of our sample, by searching within the radius corresponding to the pixel size of the extinction map (2\arcmin). The estimated A$_{Vmap}$ values are listed in the last column of Table~\ref{fit_param}. We find significant difference in A$_{V}$ from models and A$_{Vmap}$ from the extinction map (mean difference\,$\sim$\,9.3 mag), suggesting that the models need to be improved. A similar disagreement between A$_{V}$ from \cite{2007ApJS..169..328R} models and A$_{V}$ from \cite{Schultheis2009} extinction map was estimated by An11 for their spectroscopically identified YSOs.

\begin{table*}[hbt!]
\begin{center}
\begin{threeparttable}[b]
\caption{Details of the SED fit parameters. For each source represented by their SST GC No., average values of parameters from the model set corresponding to the best fit SED with the lowest $\chi_{\rm best}^{2}$ value are chosen and listed. The errors or uncertainties for parameters are roughly determined from the standard deviation in mass from all fits. The last column shows the A$_{V}$ values from the extinction map of \cite{Schultheis2009} within 2\arcmin\ of each source.} 
\label{fit_param}
\vspace{0.2cm}
\begin{tabular}{c c c c c c c c c c}
\hline
\hline 
SST GC No. & Model & N$_{data}$ & N$_{fits}$ & $\chi^{2}_{\rm best}$ & <A$_{V}$> (mag) & <Log$_{10}$(L$_{\star}$)> (L$_{\sun}$) & <T$_{\rm eff}$> (K)& <M$_{\star}$> (M$_{\sun}$) & A$_{Vmap}$ (mag)\\ 

\hline
238110 & spubhmi\tnote{a} & 8 & 2 & 96.7 & 20.5 $\pm$ 0.4 & 4.4 $\pm$ 0.0 & 8329 $\pm$ 936 & 20.0 $\pm$ 5.0 & 25.5 $\pm$ 1.1\\ 
 
352034 & sp--s-i\tnote{b} & 7 & 329 & 1.1 & 22.8 $\pm$ 2.4 & 3.7 $\pm$ 0.4 & 6668 $\pm$ 2213 & 12.4 $\pm$ 3.8 & 45.0 $\pm$ 11.3\\ 
 
358063 & sp--s-i\tnote{b} & 8 & 45 & 30.1 & 20.3 $\pm$ 0.3 & 3.7 $\pm$ 0.1 & 9091 $\pm$ 979 & 9.4 $\pm$ 1.5 & 32.3 $\pm$ 5.8\\

373107 & sp--h-i\tnote{c} & 8 & 9 & 104.5 & 24.8 $\pm$ 0.6 & 4.1 $\pm$ 0.0 & 10524 $\pm$ 499 & 15.0\tnote{*} & 31.7 $\pm$ 5.1\\  
 
388790 & spubsmi\tnote{d} & 7 & 13 & 0.9 & 22.2 $\pm$ 1.2 & 3.7 $\pm$ 0.2 & 6946 $\pm$ 1255 & 12.7 $\pm$ 2.9 & 43.0 $\pm$ 12.4\\ 

395315 & spubsmi\tnote{d} & 5 & 91 & 1.9 & 24.2 $\pm$ 3.5 & 3.2 $\pm$ 0.2 & 9397 $\pm$ 3981 & 7.8 $\pm$ 2.0 & 34.3 $\pm$ 4.2\\ 
 
405235 & spu-smi\tnote{e} & 7 & 1 & 32.7 & 20.0 & 3.9 & 6671 & 15.0\tnote{+} & 37.5 $\pm$ 2.5\\ 

609669 & spubsmi\tnote{d} & 7 & 132 & 8.6 & 22.8 $\pm$ 2.7 & 3.3 $\pm$ 0.2 & 9932 $\pm$ 3436 & 8.2 $\pm$ 1.7 & 26.7 $\pm$ 2.6\\ 
\hline 
\hline
\end{tabular} 
\begin{tablenotes}
     \item[a] A complex model in which the central star with a disk, a variable disk inner radius and bipolar cavities is enclosed in a rotationally flattened envelope structure surrounded by ambient interstellar medium.    
     \item[b] Disks around a central star, with non-variable inner radius. No surrounding envelope or ambient interstellar medium 
     \item[c] Same as b except that the disk inner radius is variable. 
     \item[d] Same as a except that the disk inner radius is set to the dust sublimation radius.
     \item[e] Same as d except that there are no bipolar cavities   
     \item[*] Only one closest stellar track to source position (see Figure~\ref{LT}) from all SED fits of the chosen model set 
     \item[+] Only one SED fit to the observed photometry by the chosen model set
   \end{tablenotes}
\end{threeparttable}
\end{center}
 \end{table*}
 

\section{Star formation rate in the CMZ} 
 
One of the commonly used ways to estimate the star formation rate (SFR) in the CMZ is by YSO counting. In this method, masses of photometrically or spectroscopically confirmed YSOs in the region are estimated either from SED fits or from zero-age main sequence (ZAMS) luminosity-mass relation. By assuming an appropriate initial mass function (IMF) and extrapolating the stellar IMF down to lower masses, the total embedded stellar population mass of the region can then be estimated. Due to the low number of spectroscopically identified YSOs, it is not possible to apply this method to our spectroscopic sample.

 However it is possible to use our photometric selection criterion (see Section~\ref{hkh8}) based on the H-K$_{\rm S}$ vs H-[8.0] diagram to obtain a much more complete sample of YSOs in the CMZ. For this, we used the the photometric catalogue of SIRIUS towards the Galactic center from which we selected our observed sample. We combine this sample with 3.6 - 8.0\,$\mu$m photometry from \cite{Ramirez2008}, 24\,$\mu$m photometry from \cite{2015AJ....149...64G} and 15\,$\mu$m photometry from ISOGAL PSC. Within $|$l$|$ $<$ 1\fdg5 and $|$b$|$<0\fdg5 we find 16,180 sources with valid photometric magnitudes in H, K$_{\rm S}$, 8.0\,$\mu$m bands and in either of the two bands: 15\,$\mu$m  or 24\,$\mu$m. We then apply our criterion (see Equation~\ref{eq1}), identifying 334 sources as YSOs. As seen in Section~\ref{hkh8}, foreground sources are removed by default using this criterion. However, OB supergiants can still contaminate our YSO sample. 
 
\begin{figure*}[!htbp]
 \centering
 {\includegraphics[width=1.0\textwidth]{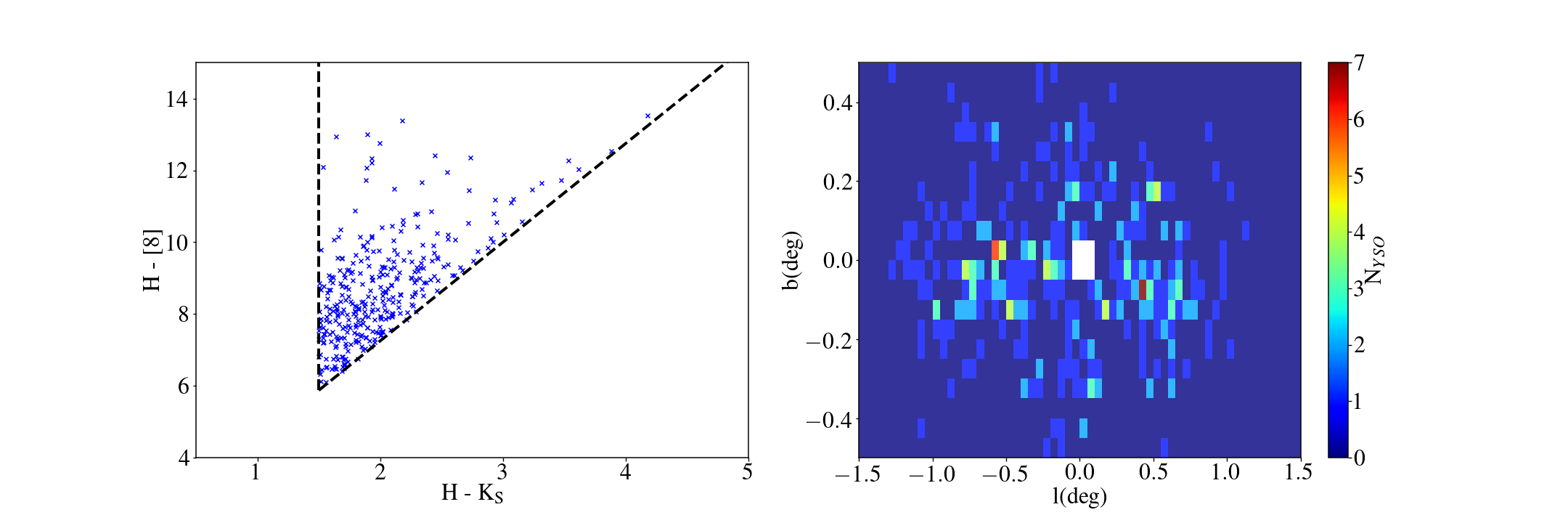}}
		\caption{Left : H-K$_{\rm S}$ vs H-[8.0] diagram used to select YSOs from the source sample constructed by combining NIR catalogue of Galactic center using SIRIUS (\citealt{1999sf99.proc..397N,2003SPIE.4841..459N}), 3.6 - 8.0\,$\mu$m photometry from \cite{Ramirez2008}, 24\,$\mu$m photometry from \cite{2015AJ....149...64G} and 15\,$\mu$m photometry from ISOGAL PSC. The dashed line represents our proposed criteria to separate YSOs and late-type stars. Right : Spatial distribution of YSOs in (l,b) plane colour coded with the number of YSOs in (l,b) bins of 0\fdg05 each. The white patch close to (l,b) $\sim$ (0\fdg0,0\fdg0) is the observational artefact from the \cite{2015AJ....149...64G} and the ISOGAL PSC catalogues where data are lacking. }
\label{NGI_YSOs}
\end{figure*}

 Figure~\ref{NGI_YSOs} shows the selected YSOs in H-K$_{\rm S}$ vs H-[8.0] diagram (left panel) and spatial distribution of YSOs in (l,b) plane colour coded with the number of YSOs in (l,b) bins of 0\fdg05 each. The white patch close to $\sim$0\fdg0 latitude and longitude is an observational artefact from the \cite{2015AJ....149...64G} and the ISOGAL PSC catalogues where data are lacking. As a result, the YSO count is higher at the negative longitudes than in positive longitudes, contrary to the fact that two-third of molecular gas is on positive longitudes (\citealt{1977ApJ...216..381B,1988ApJ...324..223B,morris1996,oka2005} etc.). We perform the SED fitting using R17 models for these 334 sources, and we determine approximate masses for them as mentioned in section~\ref{mass}. 
 
\begin{figure}[!htbp]
\centering
	{\includegraphics[width=0.49\textwidth,angle=0]{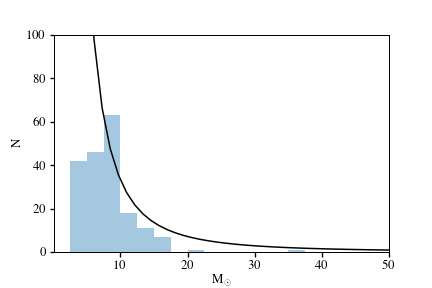}}
		\caption{The mass distribution histogram (filled blue bars) of YSOs selected using our new photometric colour-colour criteria from the SIRIUS catalog within $|$l$|$ $<$ 1\fdg5 and $|$b$|$<0\fdg5. Masses are determined approximately using the YSO parameters from SED fits and pre-main sequence tracks of \cite{1996A&A...307..829B}. The black curve represents the Kroupa IMF \citep{2001MNRAS.322..231K} fitted to the peak of the distribution. We estimate the mass of the underlying YSO population from the area under the curve from 0.01 M$_{\sun}$ to 120 M$_{\sun}$}
\label{IMF}
\end{figure}

We choose $\sim$\,190 sources with $\chi_{\rm best}^{2}$\,$<$\,35 (chosen based on the average value of $\chi_{\rm best}^{2}$ among the 8 spectroscopically confirmed YSOs) to plot the mass distribution, which ranges from 2.7\,M$_{\sun}$ to 35\,M$_{\sun}$. The distribution peaks at $\sim$\,8\,M$_{\sun}$, emphasizing that the majority of YSOs are in the high mass range. Thus we miss the low mass stars, and hence we adopt the Kroupa IMF \citep{2001MNRAS.322..231K} to fit it to the peak of our distribution extrapolate it to lower masses and estimate the total embedded stellar population in the CMZ. The Kroupa IMF for different mass ranges is given below :

\begin{equation}\label{A}
\zeta (M) = A M^{-2.3} \ \text{for} \ 0.5 M_{\sun} \leq M \leq 120 M_{\sun}
\end{equation}
\begin{equation}\label{B}
\zeta (M) = B M^{-1.3} \ \text{for} \ 0.08 M_{\sun} \leq M \leq 0.5 M_{\sun}
\end{equation}
\begin{equation}\label{C}
\zeta (M) = C M^{-0.3} \ \text{for} \ 0.01 M_{\sun} \leq M \leq 0.08 M_{\sun}
\end{equation}
\\
where A, B and C are scaling factors. We follow the method described in \cite{Immer2012} and fit our mass distribution histogram with a curve of the form as in Equation~\ref{A} by non-linear least square fitting routine. The fitting results in a value of A = 7339, which we use to obtain $\zeta (M)$ at M = 0.5 M$_{\sun}$ assuming a continuous IMF and thus estimate B = 14677 from Equation~\ref{B}. We carry out the same exercise to estimate C = 183464. Figure~\ref{IMF} shows the mass distribution histogram and the fit we performed on the distribution. Finally we estimate the total mass of YSOs to be $\sim$ 35000 M$_{\sun}$ in the CMZ using 

\begin{equation}
M_{tot} = \int_{0.01}^{120} M \zeta (M) dM  
\end{equation}

\begin{table*}[hbt!]
\begin{center}
\begin{threeparttable}[b]
\caption{Details of the SFR estimated using different methods in the literature. The method used to estimate the SFR, the region of the CMZ covered, estimated SFR and corresponding references are listed. }
\label{compare_lit}
\vspace{0.2cm}
\begin{tabular}{c c c c }
\hline
\hline 
Method & Region covered & SFR (M$_{\sun}$yr$^{-1}$) & Reference \\ 
\hline
YSO counting (photometric criterion) &  $|l|$ $<$ 1\fdg3, $|b|$ $<$ 0\fdg17 & 0.14\tnote{a} & YHA09 \\ 
\\ 
YSO counting (spectroscopic criterion) & ... & 0.07\tnote{a} &  An11 \\ 
\\
YSO counting (photometric criterion) & $|l|$ $<$ 1\fdg5, $|b|$ $<$ 0\fdg5 & 0.08\tnote{b} & \cite{Immer2012} \\ 
\\ 
\multirow{2}{*}{Free--free emission -- SFR} & $|l|$ $<$ 1\fdg0, $|b|$ $<$ 0\fdg5 & 0.015 & \cite{2013MNRAS.429..987L} \\ 
& $|l|$ $<$ 1\fdg0, $|b|$ $<$ 1\fdg0 & 0.06 & \cite{2013MNRAS.429..987L} \\
\\
Column density threshold & $|l|$ $<$ 1\fdg0, $|b|$ $<$ 0\fdg5 & 0.78 & \cite{2013MNRAS.429..987L} \\
\\
Volumetric SF relations & $|l|$ $<$ 1\fdg0, $|b|$ $<$ 0\fdg5 & 0.41 & \cite{2013MNRAS.429..987L} \\
\\
Infrared luminosity--SFR & $|l|$ $<$ 1\fdg0, $|b|$ $<$ 0\fdg5 & 0.09$\pm$0.02 & \cite{2017MNRAS.469.2263B}  \\
\\
YSO counting &  $|$l$|$ $<$ 1\fdg5, $|$b$|$<0\fdg5 & 0.046\,$\pm$\,0.026 & This work \\

\hline 
\hline
\end{tabular} 
\begin{tablenotes}
     \item[a] Assumed age of YSOs $\sim$ 0.1 Myr    
     \item[b] Assumed age of YSOs $\sim$ 1 Myr
   \end{tablenotes}
\end{threeparttable}
\end{center}
 \end{table*}

 Assuming all YSOs that constitute our sample have an average age of 0.75 $\pm$ 0.25\,Myr, we estimate the average SFR to be $\sim$\,0.046\,M$_{\sun}$yr$^{-1}$. If we assume a different IMF (e.g. Salpeter) as well as change the integration limits in the mass range in addition to including the mass uncertainties from the individual SED fitting (see Table~\ref{fit_param}) and uncertainty in the assumed age, our estimated error in the derived SFR is in the order of $\pm$ 0.026\,M$_{\sun}$yr$^{-1}$. We also changed our colour criterion by reducing the H-K$_{\rm S}$ cut to 1.0 instead of 1.5 in order to account for the variability in extinction across CMZ and the estimated SFR is still within the uncertainty limit of +0.026\,M$_{\sun}$yr$^{-1}$.
 
Our SFR estimate is lower than values from previous studies of YHA09, An11 and \cite{Immer2012}. YHA09 and \cite{Immer2012} applied YSO counting method of photometrically identified YSOs to calculate SFR of $\sim$\,0.14\,M$_{\sun}$yr$^{-1}$ (YSO lifetime $\sim$\,0.1 Myr) and $\sim$\,0.08\,M$_{\sun}$yr$^{-1}$ (YSO lifetime $\sim$\,1 Myr) respectively. An11 carried out a spectroscopic identification of YSOs among sources in common with YHA09 and derived a value of 0.07\,M$_{\sun}$yr$^{-1}$ based on the 50$\%$ contamination they found. Based on the re-examination of YHA09 sample using radiative transfer models and realistic synthetic observations, \cite{2015ApJ...799...53K} estimate the SFR to be lower by a factor of three or more. 

In addition to the YSO counting method, there have been studies that have employed the infrared luminosity-SFR relation, free-free emission from the ionised gas (i.e. bremsstrahlung radiation) at cm-wavelengths to estimate the mass of the underlying YSO population, column density threshold and volumetric star forming relations to estimate and predict the SFR in the CMZ (\citealt{2013MNRAS.429..987L,2017MNRAS.469.2263B}). \cite{2013MNRAS.429..987L} estimated the SFR in the $|$l$|$<1\fdg0, $|$b$|$<0\fdg5 to be $\sim$\,0.015\,M$_{\sun}$yr$^{-1}$ based on the free–free emission contribution to the 33 GHz flux using Wilkinson Microwave Anisotropy Probe (WMAP) data. But the predictions from the column density threshold and volumetric star formation relations exceed the observed SFR with estimates of 0.78\,M$_{\sun}$yr$^{-1}$ and 0.41\,M$_{\sun}$yr$^{-1}$ respectively. Given that predictions from these star formation relations/models are largely dependent on the mass of dense gas, it is important to make sure that different tracers of the dense gas are reliable probes. For e.g., \cite{2017ApJ...835...76M} have shown that HNCO might be a better cloud mass probe than HCN 1-0 in the Galactic center environment.

 \cite{2017MNRAS.469.2263B} found an average global SFR of $\sim$ 0.09$\pm$0.02 M$_{\sun}$yr$^{-1}$ in the same l, b range from the luminosity-SFR relations using 24\,$\mu$m, 70\,$\mu$m and total infrared bolometric luminosity. Based on the observational evidence that the individual clouds and clusters are connected along a coherent velocity structure in position-position-velocity (PPV) space \citep{2016MNRAS.457.2675H}, \cite{2017MNRAS.469.2263B} determined the SFR of individual clouds in the CMZ using the dynamical orbit model of \cite{2015MNRAS.447.1059K} assuming that star formation within these clouds is tidally triggered at the pericentre of the orbit \citep{2013MNRAS.433L..15L}. They find the total SFR within these clouds to be 0.03 to 0.071\,M$_{\sun}$yr$^{-1}$. Table~\ref{compare_lit} lists the details of the SFR estimated in the CMZ using different methods based on past studies.

Thus the SFR estimate in the CMZ from different methods (including our estimate) all point to a lower value than expected given the large reservoir of dense gas available. There are several physical explanations attributed to this dearth of star formation in the CMZ. \cite{2013MNRAS.429..987L} suggested that the additional turbulent energy in the gas, as indicated by the larger internal cloud velocity dispersion, could be providing support against gravitational collapse. The other explanations include episodic star formation in the CMZ due to spiral instabilities, high turbulent pressure in the CMZ, and the gas being not self-gravitating as discussed in detail in \cite{2014MNRAS.440.3370K}.

\section{Summary}

With the aim of estimating the SFR in the CMZ using spectroscopic identification of YSOs, we prepared a detailed observation of 22 fields using KMOS. From the 8 fields we observed, we extracted clean spectra for 91 sources. Based on the CO absorption found in cool, late-type stars and Br$\gamma$ emission seen in YSOs, we were able to clearly separate YSOs from cool, late-type stars in the EW(CO) vs EW($\rm Br {\gamma}$) diagram. We plotted our spectroscopically classified YSOs and late-type stars in the colour-colour and colour-magnitude diagrams used in the literature to classify YSOs. We found that different criteria used to classify YSOs in such diagrams were  not able to remove contaminants. We suggest a new criterion in the H-K$_{\rm S}$ vs H-[8.0] colour-colour diagram wherein we see a clear separation of YSOs and late-type stars. 

We used the new and improved version of SED models for YSOs in R17 to fit the observed photometry in the wavelength range of 1.25--24 $\mu$m for 8 YSOs. From the radii and temperatures we obtained from the SED fit, we estimated their masses to be greater than 8 M$_{\sun}$. Since we needed a bigger sample to estimate the SFR in the CMZ, we searched for sources within $|$l$|$ $<$ 1\fdg5 and $|$b$|$ $<$ 0\fdg5 with valid photometry in H, K$_{\rm S}$ (IRSF catalogue), 8.0\,$\mu$m \citep{Ramirez2008} and 15$\mu$m (ISOGAL PSC) or 24$\mu$m \citep{2015AJ....149...64G} bands. We identified 334 YSOs based on the criterion we defined in the H-K$_{\rm S}$ vs H-[8.0] diagram. We performed SED fits for these sources using R17 models resulting in ~190 sources with a good fit, and their estimated masses range from 2.7 to 35\,M$_{\sun}$, peaking at $\sim$\,8\,M$_{\sun}$. The total mass of YSOs in the CMZ was then estimated to be $\sim$\,35000\,M$_{\sun}$ by extrapolating to lower masses using a Kroupa IMF between 0.01 and 120\,M$_{\sun}$. Assuming an average age of 0.75 $\pm$ 0.25\,Myr for YSOs, we estimate the SFR to be $\sim$\,0.046\,$\pm$\,0.026\,M$_{\sun}$yr$^{-1}$, that is slightly lower than found in previous studies.

It is necessary to carry out follow up spectroscopic infrared observations to obtain a statistically significant YSO sample to further constrain our colour-colour criterion to identify YSOs. This will help in accurate determination of SFR, which is an important ingredient in the chemical evolution models of the Galaxy as well as to understand star formation of the Galactic center and as a template for circumnuclear star formation in the other galactic nuclei.

\begin{acknowledgements}
We wish to thank the anonymous referee for the extremely useful comments that greatly improved the quality of this paper. G.N and M.S. acknowledges the Programme National de Cosmologie et Galaxies (PNCG) of CNRS/INSU, France, for financial support. The research leading to these results has received funding from the European Research Council under the European Union's Seventh Framework Programme (FP7/2007-2013)/ERC grant agreement n$^{o}$ [614922].
This research made use of the SIMBAD database (operated at CDS, Strasbourg, France).
\end{acknowledgements}

\bibliographystyle{aa}
\bibliography{paper}

\end{document}